\pdfoutput=1
\documentclass[prb,twocolumn,showpacs,preprintnumbers,amsmath,amssymb,superscriptaddress]{revtex4}
 
\usepackage{graphicx}
\usepackage{dcolumn}
\usepackage{bm}
\usepackage{upgreek}

\begin{document}


\title{Elementary excitations in charge-tunable InGaAs quantum dots\\ studied by resonant Raman and resonant photoluminescence spectroscopy}
 
\author{Tim K\"{o}ppen}
\email{tkoeppen@physnet.uni-hamburg.de}
\author{Dennis Franz}
\affiliation{%
Institut f\"{u}r Angewandte Physik und Zentrum f\"{u}r
Mikrostrukturforschung, Universit\"{a}t Hamburg, Jungiusstra\ss e 11,
20355 Hamburg, Germany
}%
\author{Andreas Schramm}%
\affiliation{Optoelectronics Research Centre, Tampere University of
Technology, Korkeakoulunkatu 3, 33720 Tampere, Finland}
\author{Christian Heyn}%
\affiliation{%
Institut f\"{u}r Angewandte Physik und Zentrum f\"{u}r
Mikrostrukturforschung, Universit\"{a}t Hamburg, Jungiusstra\ss e 11,
20355 Hamburg, Germany
}%
\author{Johann Gutjahr}%
\affiliation{I. Institut f\"{u}r Theoretische Physik, Universit\"{a}t
Hamburg, Jungiusstra\ss e 9, 20355 Hamburg, Germany}
\author{Daniela Pfannkuche}%
\affiliation{I. Institut f\"{u}r Theoretische Physik, Universit\"{a}t
Hamburg, Jungiusstra\ss e 9, 20355 Hamburg, Germany}
\author{Detlef Heitmann}%
\affiliation{%
Institut f\"{u}r Angewandte Physik und Zentrum f\"{u}r
Mikrostrukturforschung, Universit\"{a}t Hamburg, Jungiusstra\ss e 11,
20355 Hamburg, Germany
}%
\author{Tobias Kipp}
\email{kipp@chemie.uni-hamburg.de}
\affiliation{%
Institut f\"{u}r Angewandte Physik und Zentrum f\"{u}r
Mikrostrukturforschung, Universit\"{a}t Hamburg, Jungiusstra\ss e 11,
20355 Hamburg, Germany
}%
\affiliation{%
Institut f\"{u}r Physikalische Chemie, Universit\"{a}t Hamburg, Grindelallee 117,
20146 Hamburg, Germany
}


\begin{abstract}
We report on resonant optical spectroscopy of self-assembled InGaAs quantum dots in which the number of electrons can accurately be tuned to $N=0,1,2$ by an external gate voltage. Polarization, wave vector and magnetic field dependent measurements enable us to clearly distinguish between resonant Raman and resonant photoluminescence processes. The Raman spectra for $N=1$ and 2 electrons considerably differ from each other. In particular, for $N=2$, the quantum-dot He, the spectra exhibit both singlet and triplet transitions reflecting
the elementary many-particle interaction. Also the resonant photoluminescence
spectra are significantly changing by varying the number of electrons in
the QDs. For $N=1$ we observe strong polaronic effects which are
suppressed for $N=2$.
\end{abstract}

\maketitle

\section{Introduction}
Semiconductor quantum dots which confine electrons are often
considered as artificial atoms and are discussed as key elements
for future applications in quantum information technology. By now,
photoluminescence and absorption spectroscopy even on single
neutral or charged QDs are well established (see, e.\ g.\, Refs.
\onlinecite{Landin1998,Bayer2000,Warburton2000,Findeis2001,Alen2003}).
Resonant Raman (or inelastic light) scattering is another widely
used spectroscopic technique to investigate the electronic
properties of semiconductor nanostructures.\cite{schueller2006}
Until now, resonant Raman spectroscopy has been very successfully
utilized to investigate electronic excitations in arrays of etched
modulation-doped GaAs-AlGaAs QDs.\cite{Strenz1994,Lockwood1996,Schueller1996,Schuller1998,
Garcia2005,Delgado2005,Kalliakos2008a,Singha2009,Singha2010} In the early papers
typical electron numbers per dot were about 100, the later papers
report on
QDs with only few electrons. There are not many reports
about resonant electronic Raman scattering in charged
self-assembled In(Ga)As QDs. In Ref.\ \onlinecite{Chu2000}, peaks
are assigned to spin-density excitations in QDs containing 6
electrons, in Ref.\ \onlinecite{Brocke2003}, charge-density
excitations in QDs containing 1 to 6 electrons are observed. Both
experiments exploit resonances via the $E_0+\Delta$ energy gap of
the QDs. 
In Ref.\ \onlinecite{Aslan2006} polarons in InGaAs QDs containing
about 7 electrons have been investigated.

In this paper, we report on a more detailed and extended study of
the results which are presented in a recent publication about
resonant spectroscopy on InGaAs QDs containing $N=1,2$
electrons.\cite{Koeppen2009} For $N=2$, these so-called QD-helium
atoms are model structures to investigate the most fundamental
many-particle states, the singlet and triplet states which resemble
the para- and the ortho-He states of the real He atoms. A key
ingredient of our experiments is that we have prepared, utilizing
the rapid thermal annealing technique,\cite{Malik1997} quantum dots
with fundamental excitation gaps of about 1.30 eV. This allows us to
excite resonantly near the fundamental $E_0$ energy gap of the QDs
and to achieve much stronger Raman intensities as compared to the
usual excitation at the $E_0+\Delta$ energy gap.
 We observe
resonantly excited PL emission peaks into excited singlet and
triplet states. Even more importantly, sharp and strong resonant
Raman transitions in the electron system both from the ground state
into singlet and triplet states and between excited singlet and
triplet states are detected. The assignment to distinct transitions
is made possible by applying an external magnetic field. We also
investigate the wave vector and polarization dependence of the
excitations which confirm that the peaks arise from inelastic light
scattering processes and give additional insight into selection
rules. In addition we compare our measurements with theoretical
calculations based on models from Merkt et al.\cite{merkt1991} and
Pfannkuche et al.\cite{pfannkuche1993} We have extended these
models to take account of the elliptical shape of the lateral
confinement in our QDs.
We also discuss, in extension to Ref.\
\onlinecite{Koeppen2009}, QDs containing $N=1$ electron. We observe
resonantly excited PL emission between the ground state of electrons
and holes and resonant Raman transitions from the electron ground
state into the first excited electron state. In addition transitions
between the first excited electron states of different angular
momentum are detected.
For all Raman processes in this work the Raman intensities were
found to be considerably strong because of the resonant excitation
near the fundamental $E_0$ energy gap of the QDs. Thus, our
experiments raise hope that the controlled manipulation of
electronic states via Raman transitions are possible even on the
level of \emph{single} QDs.

\section{Theoretical model}
\label{sec:Theo} 
\begin{figure}
\includegraphics{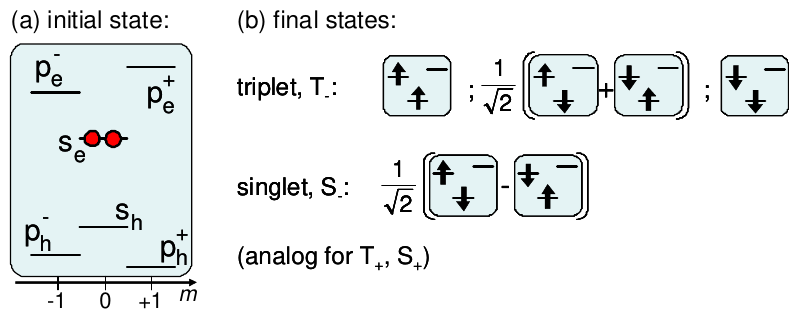}
\caption{\label{schemes}Energy level schemes: (a) Single-particle
Fock-Darwin levels for electrons (subscript e) and holes (subscript
h) in, respectively, the conduction and the valence band assuming a
parabolic confinement potential and $B>0$~T. The two electrons occupying the s$_\mathrm{e}$ shell represent the ground state of the QD helium. (b)
Representation of the QD-helium two-electron wave functions of excited singlet
and triplet states by Slater determinants. }
\end{figure}

Electronic states in In(Ga)As QDs can be described
in the simplest case by assuming a two-dimensional isotropic
parabolic potential in the lateral directions. The single-particle
energy levels in a magnetic field are then $E_{nm}=
(2n+|m|+1)\hbar\sqrt{\omega_{0_{\mathrm{e(h)}}}^2+\frac{1}{4}\omega_{c_
{\mathrm{e(h)}}}^2}+\frac{1}{2}m\hbar\omega_{c_ 
{\mathrm{e(h)}}}$.\cite{Fock1928,Darwin1930} These are the so-called Fock-Darwin
energy levels for electrons (holes) where $n$ and $m$ are the radial
and angular quantum numbers, respectively,
$\hbar\omega_{0_{\mathrm{e(h)}}}$ is the quantization energy for
electrons (holes), and $\omega_{c_
{\mathrm{e(h)}}}=\frac{eB}{m^\ast_{\mathrm{e(h)}}}$ is the cyclotron
frequency for electrons (holes), with the magnetic field $B$ and the
effective mass $m^\ast_{\mathrm{e(h)}}$ for electrons (holes). In
principle, these single-particle levels experience a Zeeman spin
splitting for $B>0 $T. The exact value of the Land\'{e} $g$ factor in
InGaAs QDs is still under discussion. Assuming a Land\'{e} $g$ factor of
2,\cite{pryor2006,pryor2007} the splitting can be estimated to be
about 0.75 meV for B=6.5 T, which is the largest magnetic field
obtainable in our experimental setup. This splitting should be
resolvable in our experiments. However, since we do not observe such
a splitting, we conclude that the $g$ factor in our QDs is smaller
than 2. In the following, we neglect Zeeman splitting and regard the
single-particle levels as two-fold degenerate due to the spin degree
of freedom. Like for atoms, shells with quantum numbers ($n,m$) of
(0,0), (0,$\pm1$), (0,$\pm2$) \ldots are labeled s, p, d, \ldots,
respectively. Figure\ \ref{schemes}(a) sketches the single particle
states for $B>0$. Here, the subscript e and h represent electrons
and holes, respectively, while the superscript gives the sign of the
angular momentum of the single particle level. This denomination is
used throughout the whole paper. In a single-particle description of
QD Helium the electronic levels are occupied by two electrons as
exemplary shown for the ground state in Fig.\ \ref{schemes}(a).

The energy levels and wave functions of QD Helium in a magnetic
field in consideration of the Coulomb interaction of both electrons
and the Pauli exclusion principle can be calculated following the
models of Merkt et al.\ \cite{merkt1991} and Pfannkuche et al.\
\cite{pfannkuche1993} The emerging two-electron wave functions can
be represented by a superposition Slater determinants. In Fig.\ \ref{schemes}(b)
such Slater determinants are sketched for the triplet and singlet
states T$_-$ and S$_-$ that contain the p$_e^-$ single-particle wave
function. Analogous representations can be sketched for T$_+$ and
S$_+$ states containing the p$_e^+$ single-particle wave function.

\begin{figure*}
\includegraphics{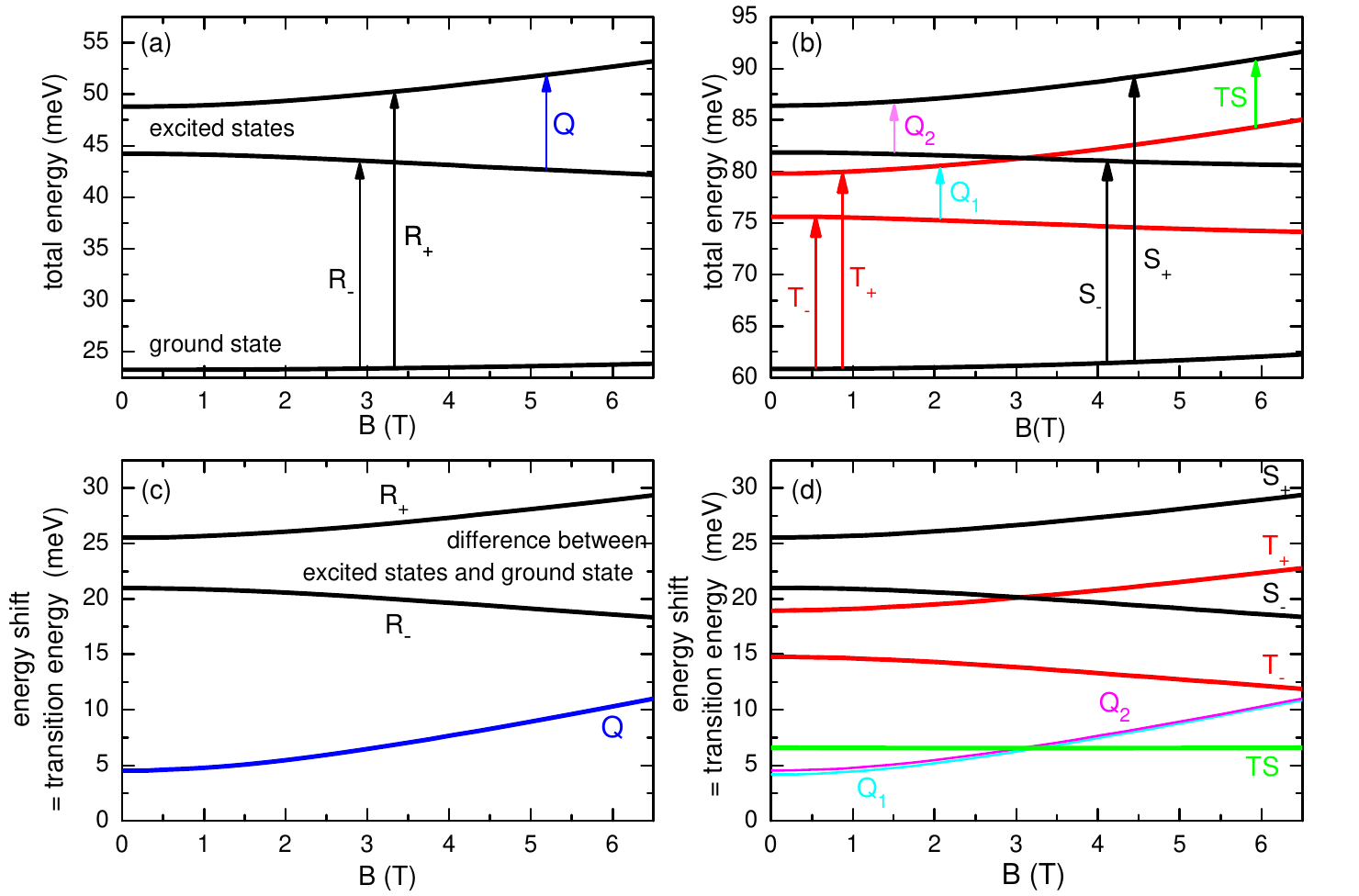}
\caption{\label{simu2}(a-b) Calculated energy levels of a QD
containing (a) $N=1$ and (b) $N=2$ electrons. For the QDs with $N=2$
both singlet (black traces) and triplet states (red traces) occur.  Transitions between these levels, as
illustrated by arrows in (a) und (b), are observed in our Raman
experiments.
(c-d) Differences between energy levels for the (c) one-electron and
(d) two-electron case.}
\end{figure*}

We extended the above models to take account of the experimentally
observed anisotropy of the lateral confinement potential assuming an
elliptical harmonic potential
$V(x,y)=\frac{1}{2}m^\ast_{\mathrm{e}}[(\omega_{0_
{\mathrm{e}}}^+x)^2+(\omega_{0_ {\mathrm{e}}}^-y)^2]$. We have
performed numerically exact calculations of the full two-electron
Hamiltonian including all many-body effects. Note that strictly
speaking, in this elliptical potential $m$ is not a good quantum
number anymore. Figures \ref{simu2}(a) and (b) show the calculated
total energy dispersion of the lowest lying states for,
respectively, $N=1$ and $N=2$ electrons. Here, we used the experimentally
observed confinement energies $\hbar\omega_{0_ {\mathrm{e}}}^+=25.6$
meV and $\hbar\omega_{0_ {\mathrm{e}}}^-=21.0$ meV that correspond to an
anisotropy splitting of $\Delta\hbar\omega=4.6$ meV. For the
one-electron case our model reproduces the analytical model of Li et
al.\cite{Li1991} For the effective mass of the electrons we have
used a value of $m^\ast_{\mathrm{e}}=0.075\,m_0$. For the $N=2$
electron case both singlet and triplet states occur. The triplet
states are at lower energies than the excited singlet states due to
the exchange energy arising from Coulomb interaction and Pauli
principle.

In Raman experiments, one measures the energy difference between a
final and an initial state. From Figs.\ \ref{simu2}(a) and (b), one
can extract transition energies by subtracting levels of total
energy from each other.  For the one-electron case the differences
between the first excited states and the ground state ($R_\pm$) as well as the
difference between the excited states of different angular momentum
(Q) are depicted in Fig.\ \ref{simu2}(c). For the two-electron
case, Fig.\ \ref{simu2}(d) depicts the transition energies from the
ground state into the excited singlet and triplet states as well as
between excited states. Here, in particular,
Q$_1=\mathrm{T}_+-\mathrm{T}_-$ and Q$_2=\mathrm{S}_+-\mathrm{S}_-$
are transitions between triplet (T$_\pm$) and between  excited
singlet (S$_\pm$) states. Note that from the calculations the
anisotropy splitting for the triplet states is slightly smaller than
for the singlet states (about 8 \% for $B=0$~T). This is the reason
why the excitation Q$_1$ is at a smaller energy shift than the Q$_2$
transition. The difference between the $Q_1$ and $Q_2$ amounts to the difference in the exchange (and correlation) energy between the two triplet states
T$_+$ and T$_-$.

In Fig.\ \ref{simu2}, only the electron states and their energy differences have been calculated. The latter are expected to be observed in Raman measurements. However, from the calculations, no prediction about the Raman intensities and, particularly, no Raman selection rules can be deduced. In order to do so, at least the interaction of the photons with the electron system has to be regarded. In the case of resonant Raman measurements, in which transitions between electron states occur via resonances involving valence-band states, also intermediate states of the electrons together with an additional electron-hole pair have to be regarded. However, such a theoretical description is beyond the scope of this paper.

\section{Sample characterization and experimental setup}
Our samples were grown by molecular beam epitaxy on a GaAs(100)
substrates. The sample we want to concentrate on in this paper has
the following layer structure: On top of the substrate a GaAs buffer
layer and a Al$_{0.3}$Ga$_{0.7}$As/GaAs superlattice were deposited.
After this, 30 nm Si-doped Al$_{0.3}$Ga$_{0.7}$As, 15 nm
Al$_{0.3}$Ga$_{0.7}$As and 40 nm GaAs were grown forming a
two-dimensional electron system (2DES) of an inverted high electron
mobility transistor. This 2DES operates as a backgate in the later
experiments. Next, one layer of self-assembled QDs were grown
exploiting the Stranski-Krastanov growth mode \cite{Leonard1994} by
depositing nominally 2.5 monolayers InAs. After a 33 nm GaAs spacer layer,
a superlattice of 16 pairs of AlAs and GaAs layers (2.5 nm each)
was grown that prevents tunneling processes of electrons to the
front gate in the later experiments. On the top of the sample, after
a 7 nm GaAs cap layer, again InAs QDs were grown with exactly the
same conditions as before. This allows us to determine the quantum
dot density by atomic force microscopy (AFM) to be about $10^{10}$
cm$^{-2}$.

To achieve that the ground state transition energy of the QDs resides in
the sensitivity range of our detector and the emission energy range of our
excitation laser, we rapidly thermally annealed the samples.\cite{Malik1997} After the rapid thermal annealing we have the possibility to excite our QDs directly with our Ti:sapphire laser (see below) at the $E_0$
energy gap. In this paper we concentrate on a sample which was annealed for 180 s at 740 $^{\circ}$C.

\begin{figure}
\includegraphics{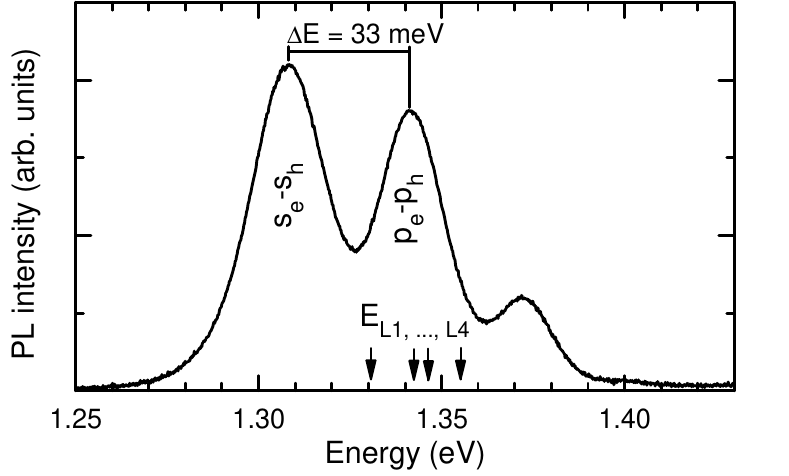}
\caption{\label{plcv}PL spectrum under nonresonant excitation
($E_\mathrm{L}=1.96$ eV). Electron-hole recombination from the s, p, and higher shells are detected. The laser energies
labeled $E_\mathrm{L1,...,L4}$ are some of the utilized energies in our
resonant measurements discussed later in the text.}
\end{figure}

We characterize our sample with nonresonant PL spectroscopy. A
typical PL spectrum is shown in Fig.\ \ref{plcv}. For these
measurements our sample was mounted in an optical cryostat and
cooled down to $T=3.3$ K. For excitation we used a HeNe laser
($\lambda=633$ nm) and for detection a Fourier transform
spectrometer with a germanium photodiode as detector. We observe at
an energy of 1.308 eV (1.108 eV before annealing) recombinations of
electrons and holes from the s states (s$_\mathrm{e}$ and s$_\mathrm{h}$) of the QDs. The next peak in the spectrum at
an energy of 1.342 eV (1.141 eV before annealing) arises from
recombinations of p-shell electrons (p$_\mathrm{e}$) and holes
(p$_\mathrm{h}$). The energy difference of these two peaks yields
that the sum of the lateral quantization energies of the electrons
and holes $\Delta E=\Delta E_\mathrm{e}+\Delta E_\mathrm{h}$ is
about 33 meV. At higher energy recombinations from higher excited
states are detected.

\begin{figure}
\includegraphics{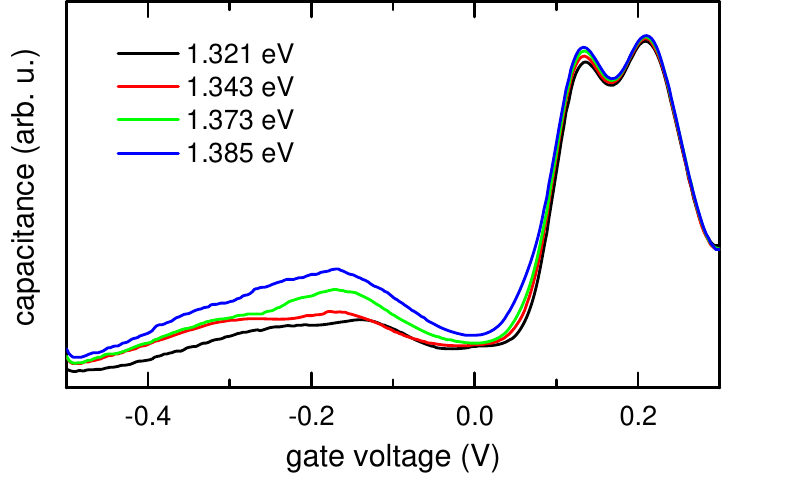}
\caption{\label{CV-Laserspot} CV spectra obtained while illuminating the sample with a focused laser (diameter about 200 $\upmu$m) of different energies $E_\mathrm{L}$, i.\ e.\, under similar experimental conditions as the later discussed Raman spectra.
}
\end{figure}

An important point of our investigations is the possibility to tune
the number of electrons in our samples. Therefore, on the top of our
sample, we have evaporated a 7 nm thick semi-transparent titanium layer that acts as a front gate. To contact the 2DES (back gate),
separate alloyed contact pads are predefined. For this purpose we
have deposited a AuGe(88:12)/Ni/AuGe layer sequence with thicknesses
of 25/5/25 nm followed by a heating to 300 $^{\circ}$C for 2
minutes. By applying a voltage between top and back gate we are able
to adjust the number of electrons in the QDs and, by measuring the
differential capacitance, to monitor the number of electrons at a
certain gate voltage. This is the so-called capacitance-voltage (CV)
spectroscopy.\cite{Drexler1994} Since in Raman experiments the
QD sample is inherently illuminated with laser light, we
investigated the charging behavior also under illumination. Figure
\ref{CV-Laserspot} shows CV spectra of the sample on which all
following Raman measurements have been performed, obtained under the
same experimental conditions as in the Raman measurements presented
below. The sample has been cooled to $T = 9$\, K and the excitation
laser of comparable intensity as in the Raman measurements (1--2~mW)
has been focused to a spot with a Gaussian full width of about 200
$\upmu$m centrally on the titanium gate on the sample. The different
spectra in Fig.\ \ref{CV-Laserspot} have been measured for different
laser energies in the range of $1.321 \textrm{ eV} \leq
E_{\textrm{L}}\leq 1.385 \textrm{ eV}$. In each spectrum two regions
can be classified: below and above $0.0$\,V gate voltage. For
positive gate voltages charging of the QDs occurs. The peaks at
about $V_{\textrm{g}}=130$\,mV and $V_{\textrm{g}}=210$\,mV, which
occur independently of the excitation laser energy, correspond to
the subsequent charging of the first and the second electron into
the QDs. The charging peaks will be analyzed in more detail below.
For negative gate voltages a photocurrent from optically excited
electron-hole pairs occurs when the internal electric field is
strong enough to overcome the exciton-binding energy and the charge
carriers can tunnel out of the QDs. This photocurrent leaves also
traces in the 90$^{\circ}$ capacitance spectrum.

In the later Raman measurements the sample is locally probed
underneath the excitation laser spot (Gaussian full width of about 200
$\upmu$m), whereas the CV spectra probe the sample underneath the
much larger titanium front gate (about 2 $\times$ 3.5 mm$^2$). To
further investigate the influence of the sample irradiation on the
QD charging, we also measured CV spectra while illuminating the gate
with a defocused laser spot of about 3 mm in diameter. We observe a
change in the photocurrent signal and, for high laser powers, an
additional photocapacitance peak occurs, \cite{Schmidt1998} however,
most importantly, the charging of the QDs occurs at about the same
gate voltages as for the focused laser.

Besides the influence of the excitation energy and the laser spot
diameter on the charging of the QDs, we have also investigated the
influence of varying excitation intensities, the magnetic field, the
position of the focused laser spot on the sample, and small changes
in the temperature. In all cases, we only observe marginal effects
on the charging of the QDs.

\begin{figure}
\includegraphics{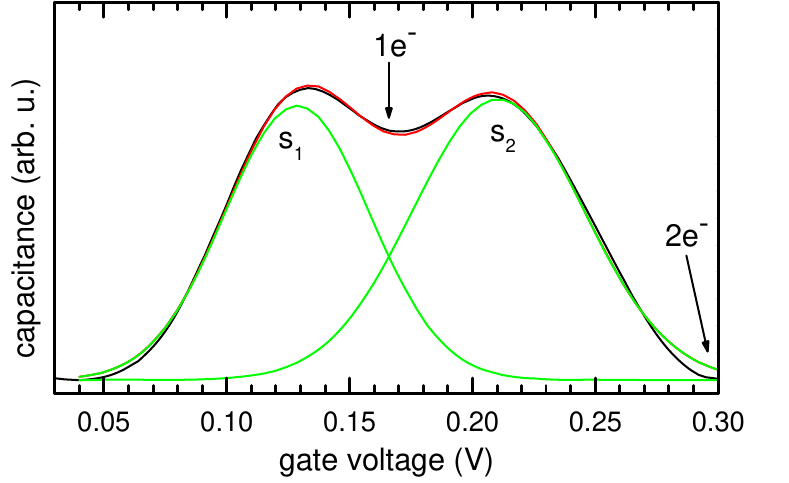}
\caption{\label{CV-Fit2} CV spectra under illumination
($E_\mathrm{L}=1.321$ eV) after subtraction of a linear
background: The charging of the s$_\mathrm{1}$ electron and the
s$_\mathrm{2}$ electron are approximated by Gaussian curves.}
\end{figure}

The most important information we want to obtain from the CV
measurements is at which gate voltage the majority of QDs are
charged with zero, one, or two electrons. Figure \ref{CV-Fit2} shows
a CV spectrum of the sample excited by a focused laser
($E_\mathrm{L} = 1.321$\,eV) after subtraction of the background.
The trace is approximated with two Gaussian curves to analyze the
charging of the QDs. The first fit represents the charging of the
QDs with the s$_\mathrm{1}$ electron and the second fit with the
s$_\mathrm{2}$ electron. From these fits, we can estimate that at a
gate voltage of $V_\mathrm{g}=160$ mV, about 85\% of the QDs
contain one electron. Consequently, our measurements of the one-electron case were performed at this gate voltage. At a gate voltage
of V$_\mathrm{g}=300$ mV almost all QDs are charged with two
electrons. Hence, the resonance and magnetic field dependent
measurements for the two electron case were executed at this gate
voltage. These measurements will be discussed in the next chapter.

For all following measurements 
we have used a Raman setup described below.
Our sample was mounted in a split-coil cryostat providing magnetic
fields up to $B=$ 6.5 T. The sample has been cooled down to a
temperature of $T=$ 9 K. For resonant excitation we have applied a
tunable Ti:sapphire laser ($850\,\mathrm{nm} < \lambda<
1000\,\mathrm{nm}$)
which was focused 
on the sample. For the spectral analysis we used a triple Raman
spectrometer with a Peltier-cooled deep-depletion CCD detector. The
spectrometer efficiently suppresses stray light and allows for
measuring as close as about 1.5 meV to the laser energy. The blazed
gratings of our spectrometer exhibit a pronounced polarization
dependence. In the spectral range of interest, $\lambda=900$ nm to
$\lambda=1000$ nm, light is diffracted about 30 times more
efficiently when polarized perpendicular to the plane of the
incidence compared to the polarization parallel to this plane. Thus,
the spectrometer acts like a polarization analyzer. For electronic
Raman spectroscopy it is fundamental to have the possibility to
adjust the polarization of the excitation laser with respect to the detected polarization of scattered
light. We have used a Fresnel rhombus to
change the polarization of the excitation laser and call a spectrum
polarized (depolarized) when the polarization of the excitation laser
is parallel (perpendicular) to the preferential polarization
direction of the detection. In the following, unless otherwise
noted, polarized spectra are shown.

\section{Experiments}
In this section we present selected spectra out of a huge number
of measurements and the evaluation of our data. First we 
concentrate on measurements at a gate voltage near
$V_\mathrm{g}=300$ mV, corresponding to $N=2$ electrons in the
QDs. Then we discuss investigations at a gate voltage near
$V_\mathrm{g}=160$ mV where most of the QDs are charged with only
$N=1$ electron.

In all following spectra the energy axes are given in the Raman
energy depiction, i.\ e., as the energy shift
$E=E_{\mathrm{L}}-E_\mathrm{det}$ between the excitation laser
energy $E_{\mathrm{L}}$ and the detection energy $E_\mathrm{det}$.
Sometimes, we combine single spectra obtained for different magnetic
fields $B$ in gray scale plots, where black (white) means high (low)
intensity. For this kind of depiction, each single spectrum was
normalized to its maximum. Dashed lines in all presented data serve
as guides to the eyes.

\subsubsection{The two-electron case}
\begin{figure}
\includegraphics{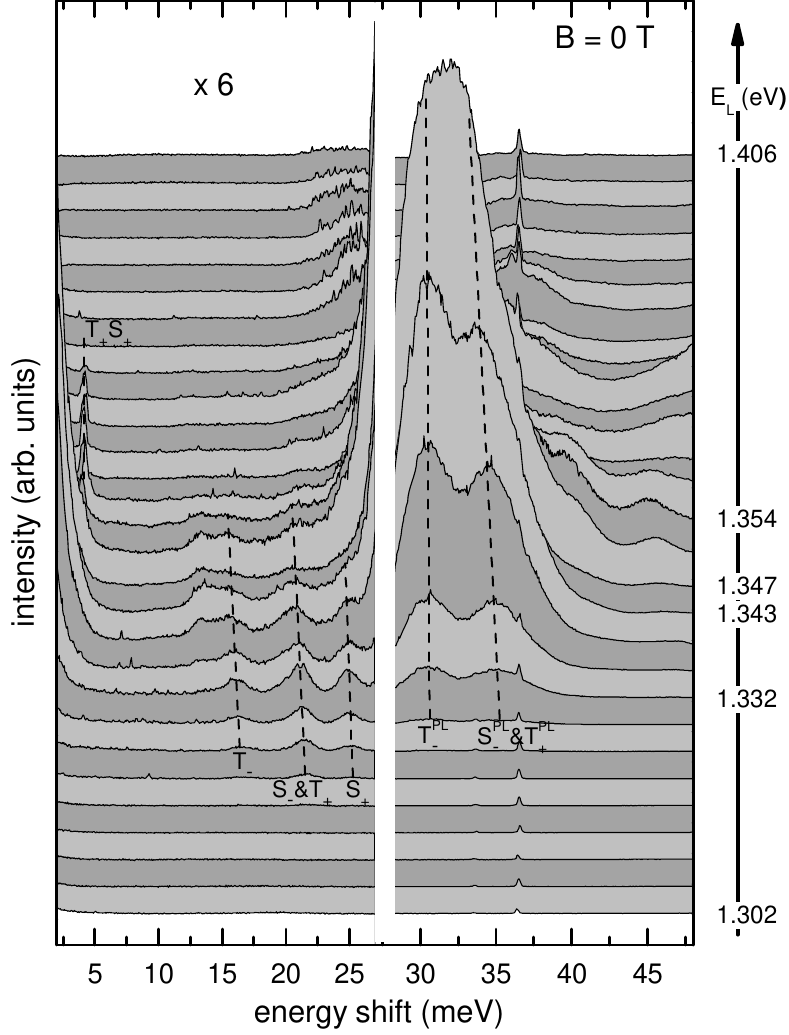}
\caption{\label{Resonanzmessung0T}Experimental spectra in the Raman
energy depiction for $B=0$ T at $V_\mathrm{g}=310$ mV corresponding
to two electrons per QD. The laser energy $E_\mathrm{L}$ is varied
between 1.302 eV and 1.406 eV in steps of $\approx3$ meV. The
spectra are vertically shifted for clarity. Below 25 meV energy shift, the  spectra
have been multiplied by a factor of 6.}
\end{figure}

\begin{figure}
\includegraphics{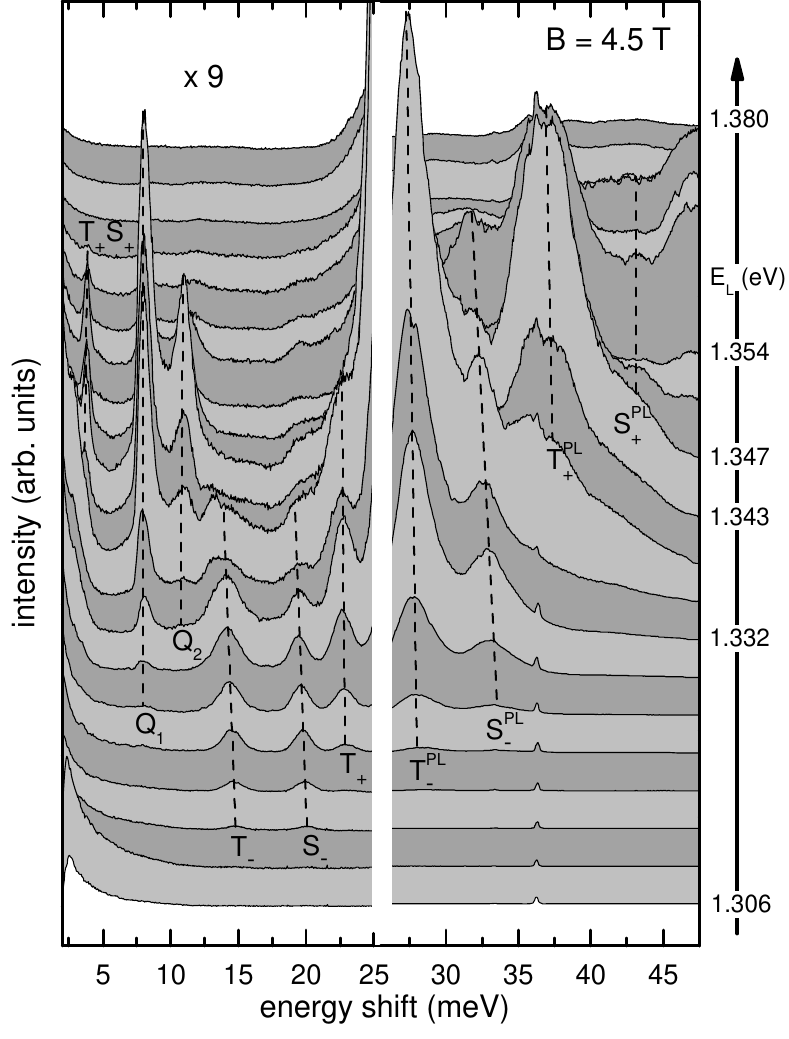}
\caption{\label{Resonanzmessung4_5T}Experimental spectra in the Raman
energy depiction for $B=4.5$ T at $V_\mathrm{g}=300$ mV
corresponding to two electrons per QD. The laser energy
$E_\mathrm{L}$ is varied between 1.306 eV and 1.380 eV in steps of
$\approx3$ meV. The spectra are vertically shifted for clarity.
Below 25 meV energy shift, the spectra have been multiplied by a factor of 9.}
\end{figure}

To get an overview of the detected excitations Fig.\
\ref{Resonanzmessung0T} shows spectra obtained for $B=0$ T and laser
energies $E_\mathrm{L}$ systematically varied between 1.302 eV and
1.406 eV in steps of $\approx3$ meV. The spectra are vertically
shifted for clarity. Intensities belonging to energies below 28 meV
have been multiplied by a factor of 6. We observe several sharp
peaks which we label with TS, T$_-$, T$_+$, S$_-$, S$_+$,
T$_-^\mathrm{PL}$, T$_+^\mathrm{PL}$, and S$_-^\mathrm{PL}$. A unique assignment of these peaks to well defined transitions is a main goal of this paper. Figure
\ref{Resonanzmessung4_5T} shows similar measurements in a magnetic
field of $B=$ 4.5 T. In these measurements we have varied the laser
energy between 1.306 eV and 1.380 eV in steps of $\approx3$ meV.
Intensities belonging to energies below 25 meV have been multiplied
by a factor of 9. The assignment of most of the peaks with labels
introduced in the previous figure was made possible by following
them in measurements with stepwise increased magnetic field. New
peaks in Fig.\ \ref{Resonanzmessung4_5T} are labeled Q$_1$, Q$_2$,
and S$_+^\mathrm{PL}$. Roughly speaking, T$_-$, T$_+$, S$_-$, and
S$_+$ become resonant for laser energies around 1.330 eV, larger
than the s$_\mathrm{h}$-s$_\mathrm{e}$- but smaller than the
p$_\mathrm{h}$-p$_\mathrm{e}$-transition energy (cf. Fig.\
\ref{plcv}). All other peaks become resonant at larger excitation
energies around 1.340 eV and 1.350 eV in the range of the
p$_\mathrm{h}$-p$_\mathrm{e}$-transition (cf. Fig.\ \ref{plcv}).
Besides the above mentioned peaks, which we will assign to
electronic transitions in the QDs, we also observe phonon
excitations at 33.6 meV and 36.6 meV, corresponding to the LO and TO
phonon of bulk GaAs, respectively.

\begin{figure}
\begin{center}
\includegraphics{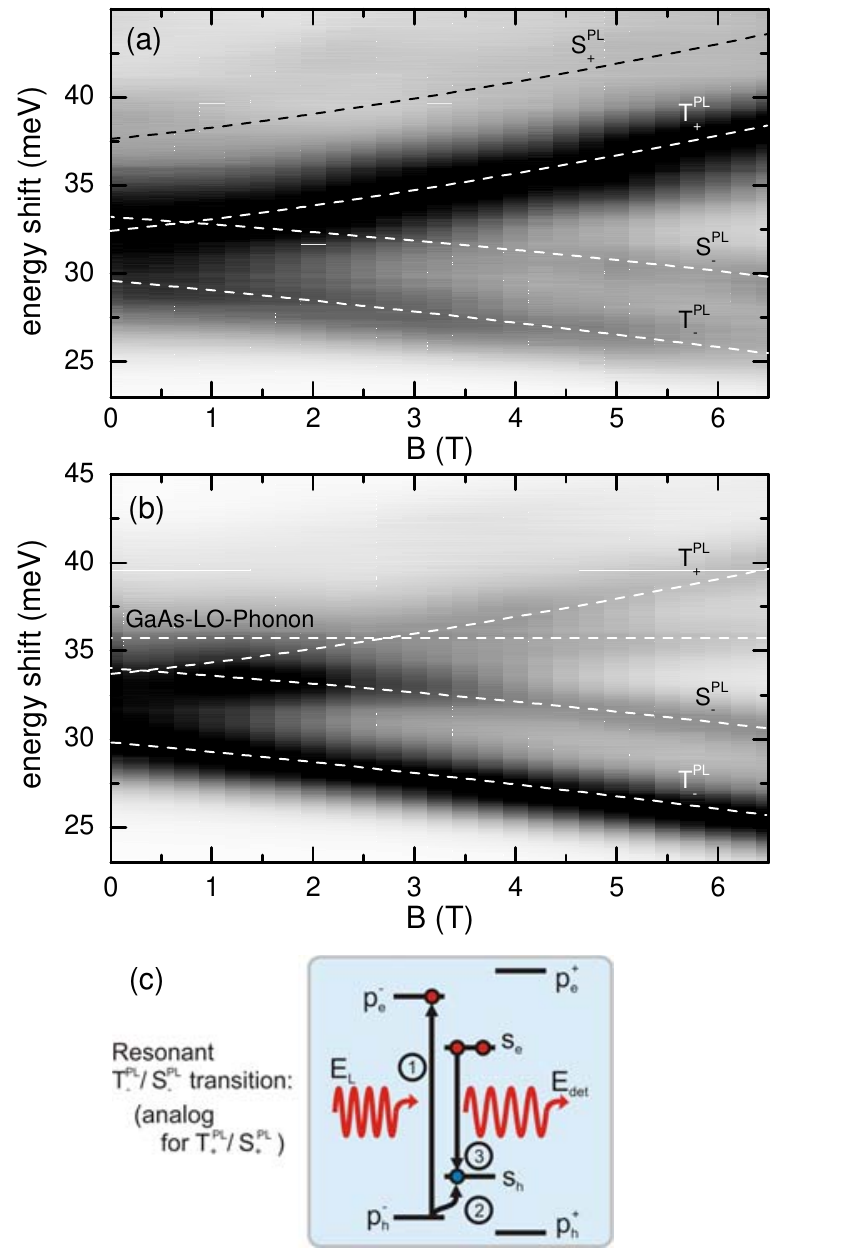}
\caption{\label{ResonantPL} (a-b) Magnetic field dispersions obtained by depicting spectra for different magnetic fields $B$ in a gray scale plot. The spectra were obtained by exciting with
laser energies of (a) $E_\mathrm{L4}=1.355$~eV and (b) $E_\mathrm{L2}=1.343$~eV.  The peaks T$_-^\mathrm{PL}$, T$_+^\mathrm{PL}$,
S$_-^\mathrm{PL}$, and S$_+^\mathrm{PL}$ arise from
resonant PL processes schematically sketched in (c) in a single-particle picture.}
\end{center}
\end{figure}

\emph{Resonant PL processes.---} In this section we concentrate on the peaks above about 25 meV which resonantly occur for excitation
laser energies larger than $E_\mathrm{L}=1.340$ eV. Figure
\ref{ResonantPL} shows the magnetic field dispersion of these peaks
when excited with (a) $E_\mathrm{L4}=1.355$ eV and (b)  $E_\mathrm{L2}=1.343$ eV,  clearly above and near the
p$_\mathrm{h}$-p$_\mathrm{e}$ transition energy (see\ Fig.\
\ref{plcv}), respectively. For the smaller excitation energy
$E_\mathrm{L2}$ the lower lying T$_-^\mathrm{PL}$ and
S$_-^\mathrm{PL}$ branches are more pronounced, whereas for the
larger excitation energy $E_\mathrm{L4}$ the higher lying
T$_+^\mathrm{PL}$ and S$_+^\mathrm{PL}$ branches are more
pronounced. The S$_+^\mathrm{PL}$ peak can hardly be seen in the
gray scale plot but it is clearly visible in the single spectra
(cf.\ Fig.\ \ref{Resonanzmessung4_5T}). We assign these branches to
resonant PL processes as sketched in the single-particle picture in
Fig.\ \ref{ResonantPL}(c). Exciting with a laser energy resonantly
matching the p$_\mathrm{h}$-p$_\mathrm{e}$ transition energy,
electron-hole pairs are created in the QDs. After the excitation,
the electron in the p shell cannot relax into the s shell since it
is already completely filled with two electrons. On the other hand,
the hole will quickly relax into its s state which comes along with
energy dissipation into the crystal lattice. Then a radiative
s$_\mathrm{e}$-s$_\mathrm{h}$ recombination process takes place
leaving the QDs behind in a configuration with one electron in the s
state and the other electron in the p state. Such configuration
forms either a singlet or a triplet state, in analogy to the para
and ortho He in real atoms.

The resonant excitation is inevitable to resolve splittings between
singlet and triplet states in the PL spectra of an inhomogeneously
broadened QD ensemble. For a particular excitation energy in the
range of the p$_\mathrm{h}$-p$_\mathrm{e}$ transition automatically
two subensembles of QDs are excited: one with the matching
p$_\mathrm{h}^-$-p$_\mathrm{e}^-$ transition energy, the other one
with the matching p$_\mathrm{h}^+$-p$_\mathrm{e}^+$ transition
energy. These subensembles differ mostly in their quantization in
growth direction and exhibit only small variations in the lateral
quantization energy. After excitation and hole relaxation, radiative
recombination within each subensemble into singlet and triplet
states occurs, explaining the observed S$_-^\mathrm{PL}$,
T$_-^\mathrm{PL}$, S$_+^\mathrm{PL}$, and T$_+^\mathrm{PL}$
branches where the indexed sign refers to the sign of $m$ of the remaining p electron in one of the two energy split p states.
The linewidths of the observed resonant PL signals are in the range
of 2 meV which is a measure of the distribution of lateral
quantization energies of the subensemles. This value is much smaller
than the linewidths of the PL peaks in the nonresonant measurements
in which virtually all QDs under the laser spot are excited
nonresonantly (cf.\ \ref{plcv}). Their width of about 22 meV is a
measure of the distribution of the total quantization energy in
lateral and in growth direction.

The T$_-^\mathrm{PL}$ and T$_+^\mathrm{PL}$ branches are not
degenerated for $B=0$ T, where we observe a splitting of about 5
meV. The lifting of degeneracy can be explained by a slightly
elliptical lateral potential
\cite{Babinski2006,Hameau1999,Carpenter2006} which primarily arises
from the lateral elliptical shape of our QDs  that is typical for
our molecular beam epitaxy growth conditions. The
influence of piezoelectric effects is small for the rapidly
thermally annealed QDs.\cite{Schliwa2007}

The strong difference in intensity between triplet and singlet
lines, which we observe, has been reported before in nonresonant PL
measurements on single QDs.\cite{Warburton2000,Findeis2001} It is
explained by the larger degeneracy of the triplet state and
electron-hole exchange interactions.

\begin{figure}
\includegraphics{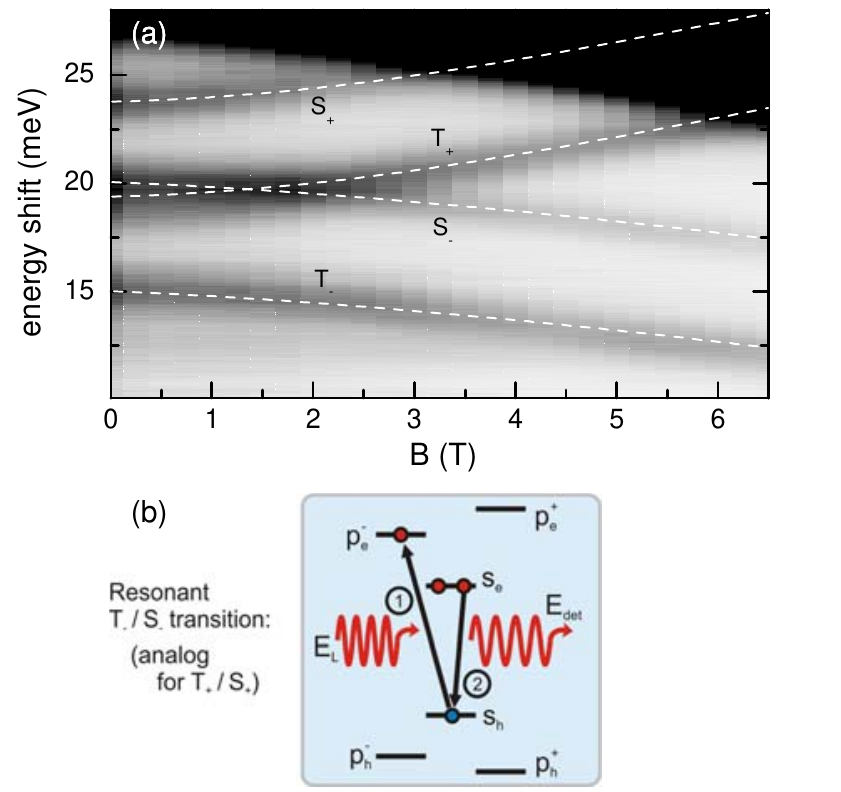}
\caption{\label{Raman194Wez}(a) Magnetic field dispersion of peaks T$_-$, T$_+$,
S$_-$, and S$_+$ excited with a
laser energy of $E_\mathrm{L1}=1.331$ eV. These peaks are assigned to resonant Raman processes. (b) Sketch of the resonant Raman processes in a single-particle picture.}
\end{figure}

\emph{Resonant Raman scattering processes.---} In this section we concentrate on the peaks that occur in the energy
range of 13 meV to 28 meV in the spectra of Figs.\
\ref{Resonanzmessung0T} and \ref{Resonanzmessung4_5T} for excitation
energies $E_\mathrm{L}$ close to 1.33 eV, i.\ e.\ below the
p$_\mathrm{h}$-p$_\mathrm{e}$ transition in the range of the
s$_\mathrm{h}$-p$_\mathrm{e}$ transition energy (see\ Fig.\
\ref{plcv}).
Figure \ref{Raman194Wez}(a) shows the magnetic field dispersion of
these peaks for $E_\mathrm{L1}=1.331$ eV. Since the intensities of
these peaks are a factor of about 50 lower than the above described
resonant PL peaks, we have changed the intensity scaling in the gray
scale plots such that the excitations in this range are clearly
visible.

The four branches labeled T$_-$, T$_+$, S$_-$, and S$_+$ arise from
Raman scattering processes as will be confirmed later by the
polarization and wave vector dependence of the excitations. We
assign the peaks to resonant Raman excitations from the ground state
into excited triplet (T$_-$, T$_+$) and singlet (S$_-$, S$_+$)
states. Exemplarily, the scattering process for the T$_-$ and S$_-$
peaks is visualized in Fig.\ \ref{Raman194Wez}(b) in a simple
single-particle picture as a two step process. In the first step an
excitation from the s$_\mathrm{h}$ to the p$_\mathrm{e}$ state of
the QDs occurs followed by a radiative recombination process between
an electron of the s$_\mathrm{e}$ shell and a hole of the
s$_\mathrm{h}$ shell in the QDs. 
Effectively, in the single-particle picture, we excite an electron
from the s$_\mathrm{e}$ state to a p$_\mathrm{e}$ state. Beyond the
single-particle picture, the s$_\mathrm{e}$ and p$_\mathrm{e}$
electrons form either an excited singlet or triplet state, as
represented by Slater determinants in Fig.\ \ref{schemes}(b). In
contrast to the resonant PL, by measuring the Raman peaks we can
extract the singlet and triplet transition energies directly without
any assumptions on the confinement energies of the holes. The
resonant Raman excitations
are observed about 13 meV below the corresponding resonant PL peaks (see
Fig.\ \ref{ResonantPL}). This value matches with the energy of the
non-radiative p$_\mathrm{h}$-s$_\mathrm{h}$ relaxation of the hole
in the resonant PL process. Thus, the difference between the PL and
Raman peaks in our experiments gives the hole quantization energy
which is in the range of a third of the total lateral confinement energy in our
measurements. This finding for the hole quantization is in good
agreement with previous theoretical and experimental work on single
In$_{0.5}$Ga$_{0.5}$As QDs.\cite{Findeis2001a} The linewidths of
the detected Raman peaks are in the range of 1.5 meV, slightly
smaller than the ones of the resonant PL peaks due to the lack of
deviations in the hole quantization energy.
\begin{figure}
\includegraphics{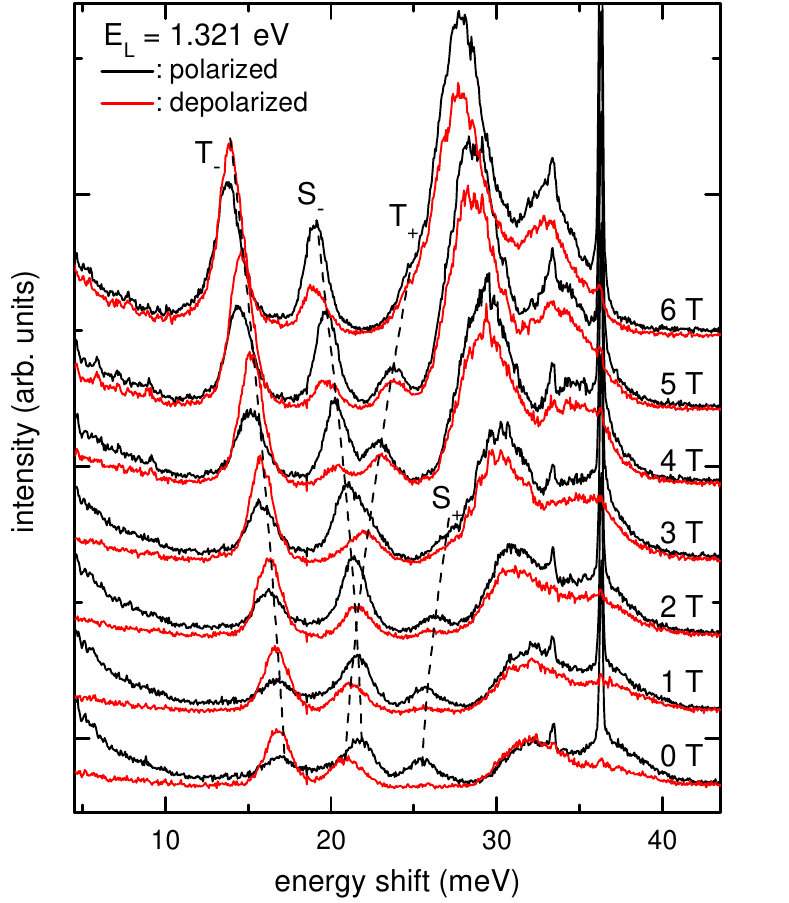}
\caption{\label{Polarisationsvergleich}Polarization dependence of
the resonant Raman and PL excitations in a magnetic field up to
$B=6$ T ($E_\mathrm{L}=1.321$ eV). Black (red) traces are
polarized (depolarized) spectra. The spectra are vertically
shifted for clarity. We observe a clear polarization dependence
for the Raman excitations which is softened by increasing the
magnetic field.}
\end{figure}

In contrast to the resonant PL peaks, the Raman peaks T$_\pm$ and
S$_\pm$ have nearly the same intensity, manifesting that indeed they
arise from a fundamentally different process than the
T$_\pm^\textrm{PL}$ and S$_\pm^\textrm{PL}$ peaks.

Furthermore, the Raman signals exhibit a clear polarization
dependency. For 2DES, Hamilton et al.\cite{Hamilton1969} have
calculated that in polarized configuration, i.\ e.\, with parallel
polarization of the excitation and detected light, collective
charge-density excitations (CDEs) can be observed. In the
depolarized case, i.\ e.\, with the polarization of the excitation
and scattered light perpendicular to each other, collective
spin-density excitations (SDEs) can be detected. These polarization
selection rules were also confirmed for deep-etched AlGaAs/GaAs QDs
by different groups.
\cite{Lockwood1996,Schueller1996,Schuller1998,Garcia2005,Kalliakos2008a,Delgado2005,Singha2009,Singha2010}
In Fig.\ \ref{Polarisationsvergleich}, polarized and depolarized
spectra for various magnetic fields are compared for our QDs charged
with $N=2$ electrons. Here, we have used an excitation laser energy
of $E_\mathrm{L}=1.321$ eV. For $B=0$ T in the polarized spectra the
excitations into the excited singlet states are observed to be more
intense whereas in the depolarized spectra the excitations into the
excited triplet states are more pronounced. Thus, the excitations
into excited singlet (triplet) states have charge (spin) density
character. This is in agreement with considerations that for
excitations from the singlet ground state into the triplet states a
spin flip is needed and this is only possible for SDEs. Note that
for the same reason also direct optical dipole excitations of the
triplet state with far-infrared light are not possible. For
field-effect-confined quantum dots charged with $N=1,...,4$
electrons, only transitions into the singlet state are observed in
far-infrared measurements.\cite{Meurer1992} The observed
polarization dependence is in accordance with the polarization
selection rules of Raman excitations in 2DES and in etched
GaAs/AlGaAs QDs. With increasing magnetic field we observe a
softening of the Raman polarization selection rules as it has been
calculated and observed for deep-etched QDs.\cite{Schuller1998,Delgado2005,Kalliakos2008a} The resonant PL
signal shows no polarization dependence at $B=0$ T but we observe
small polarization dependencies for $B>0$ T. In Fig.\
\ref{Polarisationsvergleich}, the increase of the T$_-$ and S$_-$
peak intensities and the decrease of the T$_+$ and S$_+$ peak
intensities with increasing magnetic field is because the
subensemble of QDs in resonance with the
s$_\mathrm{h}$-p$_\mathrm{e}^-$ transition gets larger whereas the
subensemble in resonance with the s$_\mathrm{h}$-p$_\mathrm{e}^+$
transition gets smaller with increasing $B$ for the particular
excitation energy $E_\mathrm{L}$. The same explanation holds for the
intensity trend of the resonant PL peaks.

\begin{figure}
\includegraphics{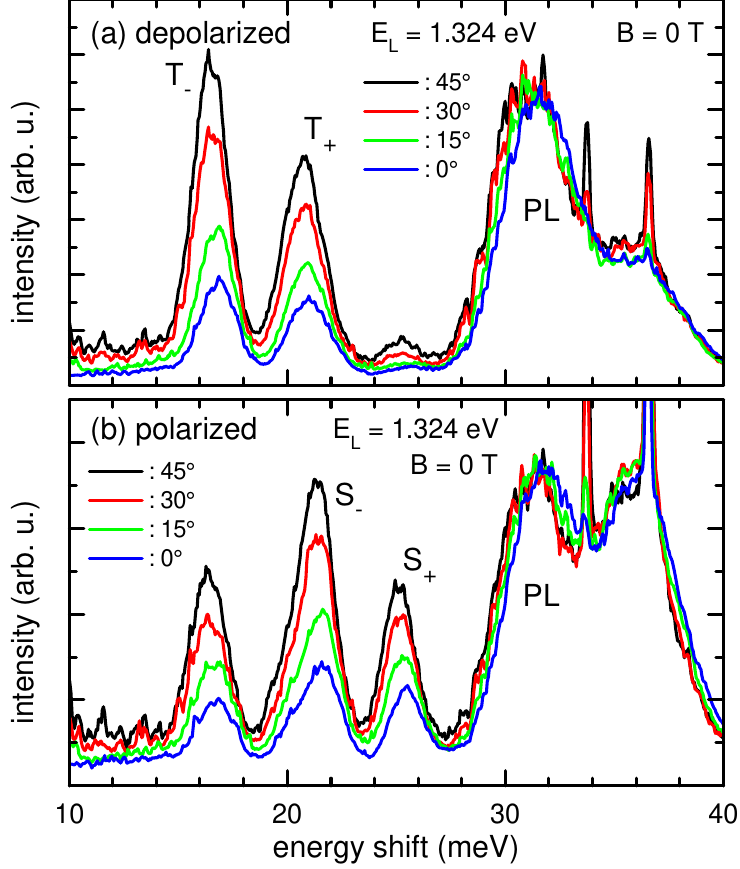}
\caption{\label{wavevector}(a) Depolarized and (b) polarized spectra
for different angles of incidence 0$^\circ$, 15$^\circ$, 30$^\circ$ and 45$^\circ$ corresponding to lateral
wave vector transfers $|\textbf{q}|$ of 0 cm$^{-1}$, $3.47\times 10^4$ cm$^{-1}$, $6.71\times 10^4$ cm$^{-1}$,
and $9.49\times 10^4$ cm$^{-1}$, respectively. The spectra are normalized to the PL peak
at 31 meV. We observe
strongly enhanced Raman peaks by increasing the wave vector
transfer.}
\end{figure}

We also investigated the influence of a lateral wave vector transfer
in our resonant optical measurements. To achieve this we have tilted
the sample with respect to the incoming laser beam. In the
back-scattering geometry, since the incident and scattered
wavelengths $\lambda$ are similar, the relation between the angle
$\theta$ measured to the sample's normal and the wave vector
transfer $|\textbf{q}|$ is given by
$|\textbf{q}|=\frac{4\pi}{\lambda}\sin\theta$. Figure
\ref{wavevector} shows (a) depolarized and (b) polarized spectra for
different tilting angles ($\theta=$ 0$^\circ$, 15$^\circ$,
30$^\circ$, and 45$^\circ$) for an excitation laser energy of
$E_\mathrm{L}=1.324$ eV. The spectra are normalized to the PL signal
at about 31 meV. First of all it is obvious that the peaks do not
show a wave vector dispersion but stay at a constant energy as one
would expect for  zero-dimensional electron systems in QDs.\cite{Schueller1999,Barranco2000} Secondly, compared to the
resonant PL peaks, the Raman peaks are strongly enhanced with an
increased wave-vector transfer. The underlying single-particle
excitation for the Raman process from the ground state into both the
excited singlet and triplet state is from s$_\mathrm{e}$ to
p$_\mathrm{e}$, thus, these excitations are dipole excitations. The
Raman process is a two photon process for which the parity is
conserved in first approximation for a symmetric system, thus the
dipole excitations should be Raman forbidden. By transferring a
lateral wave vector the symmetries are broken and the parity
selection rule is weakened, as has already been observed for etched
GaAs/AlGaAs QDs.\cite{Strenz1994,Schuller1998} This is also the
reason why in our case the resonant Raman peaks get stronger with
increasing wave vector transfer. However, the occurrence of dipole
transitions even for negligibly small $|\textbf{q}|$ proves the
parity selection rules in our QDs to be inherently weakened. The main
reason for that might be the anisotropy in the lateral potential of
the QDs.

\emph{Transitions between excited singlet and triplet states.---} In this section we concentrate on the peaks that occur in the energy
range between 3 meV and 13 meV in the spectra of Figs.\
\ref{Resonanzmessung0T} and \ref{Resonanzmessung4_5T} for excitation
energies $E_\mathrm{L}$ between 1.340 eV and 1.350 eV.

\begin{figure}
\includegraphics{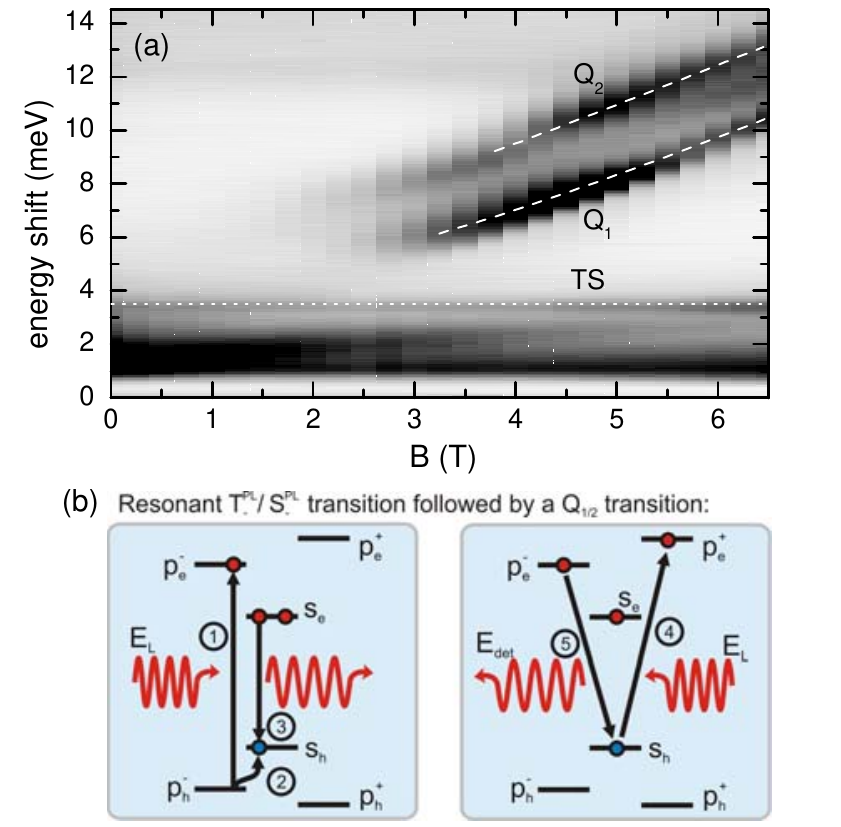}
\caption{\label{DispersionST}
(a) Magnetic field dispersion of peaks Q$_1$,Q$_2$, and TS excited with a
laser energy of $E_\mathrm{L3}=1.347$~eV. These peaks are assigned to transitions
between excited triplet/singlet (Q$_1$/Q$_2$) states of different
angular momentum and between excited triplet and singlet states
(TS). (b) Scheme of the doubly resonant excitation process for Q$_{1/2}$ in a single-particle picture. It consist of a resonant PL process (left scheme) followed by a resonant Raman process (right scheme).
}
\end{figure}

Figure \ref{DispersionST}(a) shows the magnetic field dispersion of
these peaks for $E_\mathrm{L3}=1.347$ eV, which is in the range of
the p$_\mathrm{h}$-p$_\mathrm{e}$ transition. We assign the highly
dispersive branches Q$_1$ and Q$_2$ to transitions between excited
triplet (from T$_-$ to T$_+$) or singlet (from S$_-$ to S$_+$)
states, respectively. The underlying processes are sketched in the
single-particle picture in Fig.\ \ref{DispersionST}(b). First, an
excited triplet or singlet state has to be created, which is
achieved by the resonant PL process discussed before. Then the
actual Raman process occurs as follows: In a first step a resonant
s$_\mathrm{h}$-p$_\mathrm{e}$ transition occurs. After this, a
radiative recombination between the electron from the
p$_\mathrm{e}^-$ with the hole from the s$_\mathrm{h}$ state takes
place. Thus, effectively, an electron from the p$_\mathrm{e}^-$
state is transferred into the p$_\mathrm{e}^+$ state. For this
combined process two resonance conditions have to be fulfilled: A
resonant excitation of a singlet or a triplet state via a PL process
and a resonant excitation of the Raman process. Hence, these
transitions can only be observed for certain magnetic fields, for
which the p$_\mathrm{h}^-$-p$_\mathrm{e}^-$ matches the
s$_\mathrm{h}$-p$_\mathrm{e}^+$ transition energy.
We assign the excitation Q$_1$ to a Raman scattering process between
the T$_-$ and the T$_+$ branches because its energy nicely matches
with values obtained by subtracting the energy values of the T$_-$
from the T$_+$ or the T$_-^\mathrm{PL}$ from the T$_+^\mathrm{PL}$
branch. The Q$_2$ excitation is assigned to the transition between
the excited singlet states, i.\ e.\, from the S$_-$ to the S$_+$
state. It matches with values achieved by subtracting the
S$_-^\mathrm{PL}$ from the S$_+^\mathrm{PL}$ branch. It is not
possible to extract the difference between the S$_+$ and the S$_-$
branch due to superimposed PL peaks in this range. The Q$_1$ peaks
have larger intensities than the Q$_2$ peaks (cf.\ Fig.\
\ref{Resonanzmessung4_5T}). This can be explained by a larger
degeneracy and longer relaxation lifetimes of the triplet in
comparison to the singlet state.\cite{Warburton2000,Alen2004}

The peak labeled with TS at an energy of about 4 meV
exhibits nearly no dispersion in the magnetic field. We
tentatively assign this excitation to a transition between
excited triplet and singlet states of similar dispersion. More precise, resonance measurements described below suggest that the TS branch arises from excitations from T$_+$ to S$_+$ states. The process can be described in the single-particle picture by a resonant PL process into the T$_+$ state and a subsequent two-step Raman process in which a resonant p$_\mathrm{h}^+$-p$_\mathrm{e}^+$ excitation is followed by a p$_\mathrm{e}^+$-p$_\mathrm{h}^+$ recombination leaving the QD behind in a S$_+$ state. For the excitations Q$_1$, Q$_2$,
and TS we do not detect a distinctive polarization dependency.

\begin{figure*}
\includegraphics{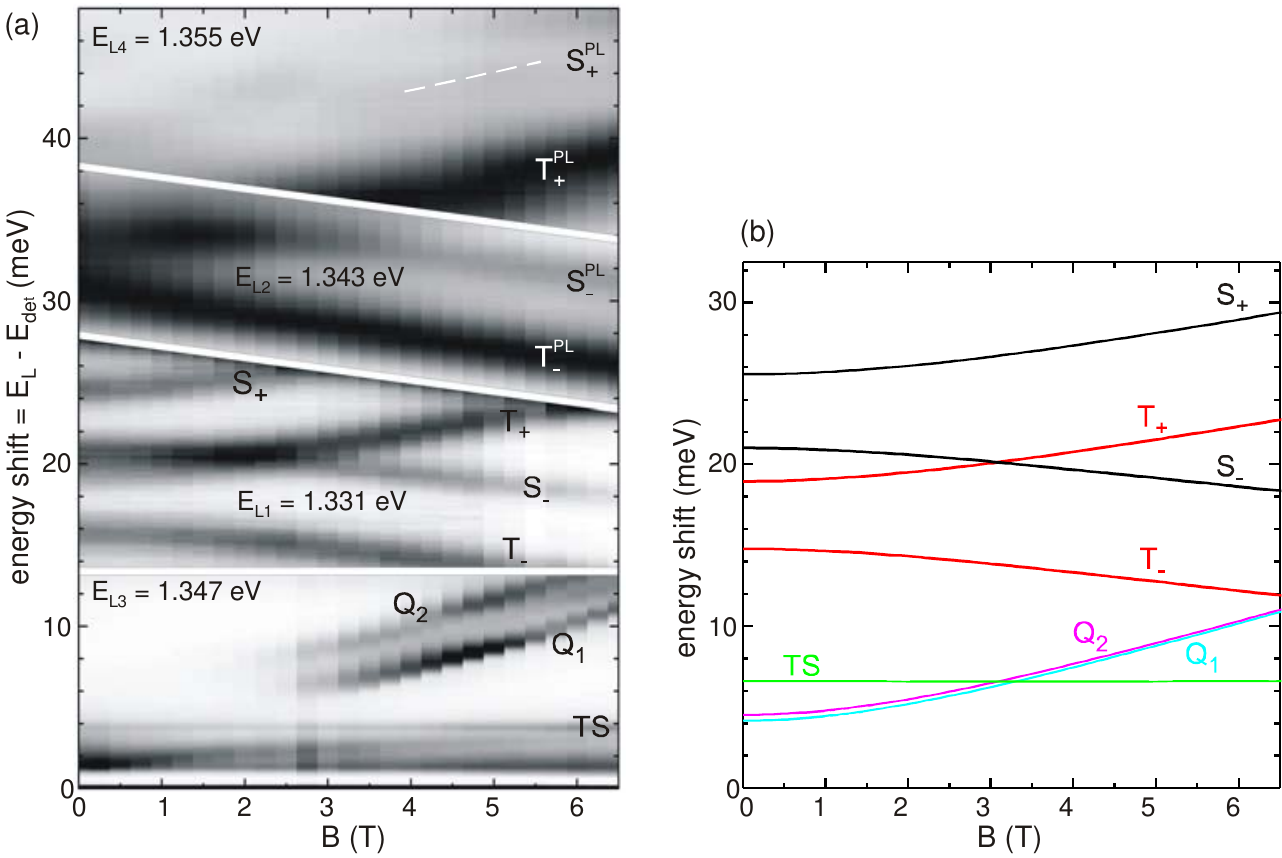}
\caption{\label{Uebersicht}Comparison between experiments and
theory: (a) Summary of the magnetic field dispersion of all observed peaks for  resonant excitation with different laser
energies $E_\mathrm{L1}$ to $E_\mathrm{L4}$. This graph is a combination of the dispersion graphs in Figs.\ \ref{ResonantPL}, \ref{Raman194Wez}, and \ref{DispersionST}. Regions of different laser
energies are separated by white gaps. (b) Calculated transition energies for QDs containing $N=2$ electrons assuming an elliptical potential in lateral direction.}
\end{figure*}

\emph{Comparison to the theoretical model.---} Figure \ref{Uebersicht} shows a compilation of the gray scale plots
of Figs.\ \ref{ResonantPL}, \ref{Raman194Wez}, and
\ref{DispersionST}. It gives a complete picture of the dispersion of
all peaks observed for the different excitation laser energies
$E_\mathrm{L1}$ to $E_\mathrm{L4}$ close to their respective
resonance. For comparison, Fig.\ \ref{Uebersicht}(b) shows  the
theoretical calculations assuming an elliptical harmonic potential
in lateral direction for QDs containing $N=2$ electrons, obtained as
described in Section \ref{sec:Theo}. From the measurements we obtain
that for $B=0$~T the transition energies into the excited triplet
states are about \mbox{78\%} of the ones for excited singlet states
at $B=0$ T. The calculations give a value of about \mbox{71\%} which
is in good agreement to the experiments.

The calculations as well reproduce the non-degeneracy of the Q
branches. In both, experiment and theory, the branch representing
transitions between excited triplet states of different angular
momentum (Q$_1$) occurs at lower energy compared to the branch
of the transitions between the excited singlet states (Q$_2$). The
non-degeneracy is peculiar for an anisotropic, in our case
elliptical harmonic potential. For an isotropic harmonic potential
these branches would be degenerate. In Fig.\ \ref{Uebersicht}, the
calculated Q$_1$ and Q$_2$ excitations are depicted in the range of
$B=0$ T to $B=6.5$ T although, experimentally, they can only be
observed in a small range of the magnetic field due to the resonance
conditions. Apart from the qualitative agreement, quantitatively the
splitting between both Q branches is larger in the experiment than
in the calculation. Also the experimental TS branch quantitatively
differs from the calculated branch.

Small deviations between measurements and calculations arise most
likely from the fact that the values for the effective mass $m^\ast$
and the dielectric constant $\epsilon$ which are needed for the
calculations are not well-known for the QDs. We have assumed for
$\epsilon$ the value of InAs 15.15 and for
$m^\ast=0.075\,m_\mathrm{e}$ which is in accordance to measurements
of Fricke et al.\ on InAs QDs.\cite{Fricke1996} Furthermore, it is
not a priori known how well the assumption of an elliptic harmonic
potential fits to the real system.

Additional deviations can occur due to an artificial dispersion
provoked by different subensembles of QDs. As mentioned above, for a
particular excitation laser energy $E_{\mathrm{L}}$, particular
subensembles for QDs are excited that fulfill the resonance
conditions. By changing the magnetic field but keeping $E_\mathrm{L}$
fixed, the resonantly excited subensembles change. An artificial
dispersion then can occur, when the different subensembles exhibit
gradually changing lateral quantization energies. A close look on
Fig.\ \ref{Resonanzmessung0T} reveals that the resonant Raman peaks
T$_\pm$ and S$_\pm$ slightly decrease their energy shift with
increasing $E_\mathrm{L}$: Increasing $E_\mathrm{L}$ by 30 meV leads
to a decrease of the energy shift of about 1.3 meV. Coming back to
the measured magnetic field dispersion, increasing $B$ from 0 T to
6.5 T leads to a shift of the resonant Raman peaks of roughly 5 meV.
Thus, the amount of the artificial dispersion included in this shift
can roughly be estimated to be $1.3/30 \times 5 \mathrm{~meV} = 0.22
\mathrm{~meV}$, i.\ e.\ negligibly small. In comparison, Preisler et
al.\  have investigated multistacks of 20 layers
of InAs QDs by near-resonant PL spectroscopy.\cite{Preisler2009} They observe energy
shifts of the resonant PL excitations that are about a factor of 3-4
larger than in our measurements, probably because the stacking of
QDs might lead to a broader size distribution of the QDs.

\begin{figure}
\includegraphics{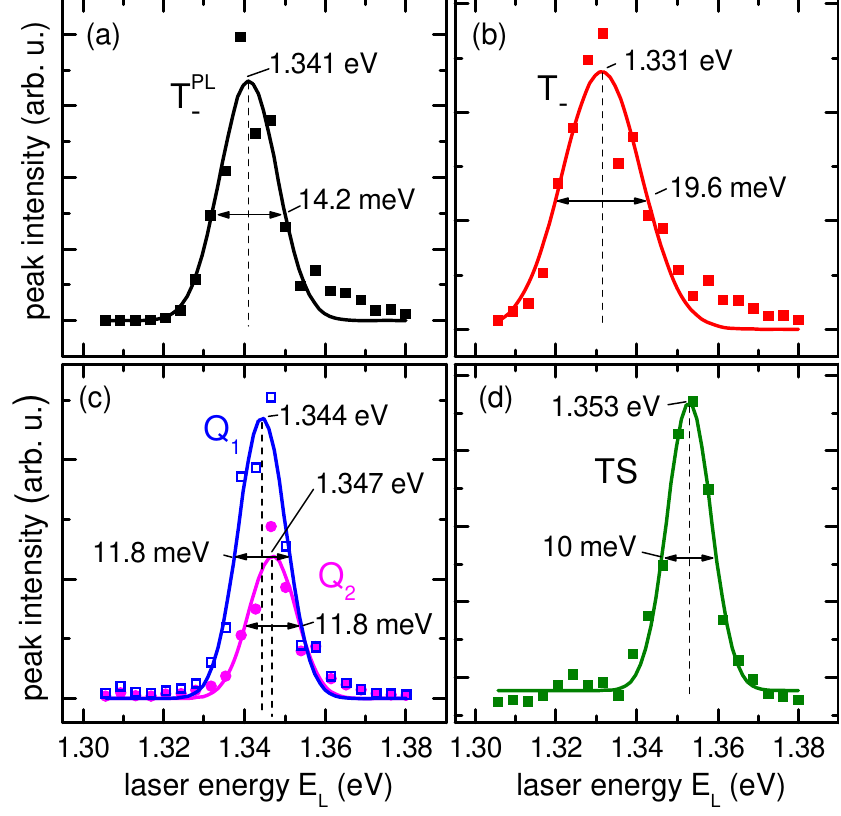}
\caption{\label{Resonanz}The peak intensity of (a) T$_-^\mathrm{PL}$, (b) T$_-$, (c) Q$_1$ and Q$_2$, and (d) TS in dependence
of the excitation laser energy E$_\mathrm{L}$ as extracted from the measurements of Fig.\ \ref{Resonanzmessung4_5T} for
(a), (b), and (c) and of Fig.\
\ref{Resonanzmessung0T} for (d). The measured data points are approximated by Gaussian profiles.
The FWHMs and the laser energies where the
excitations have their intensity maximum are given in the plots.
}
\end{figure}

\emph{Resonance behavior.---} Figure \ref{Resonanz} depicts resonance traces for the peaks (a)
T$_-^\mathrm{PL}$, (b) T$_-$, (c) Q$_1$ and  Q$_2$, each for $B=4.5$~T,
and (d) TS for $B=0$~T. These traces give the dependence of the peak
intensities on the excitation laser energy $E_\mathrm{L}$ extracted
from the spectra shown in Figs.\ \ref{Resonanzmessung4_5T} and
\ref{Resonanzmessung0T}. The data points from the measurements are
approximated by Gaussian profiles. For the T$_+$, S$_-$, S$_+$,
T$_+^\mathrm{PL}$, S$_-^\mathrm{PL}$, and S$_+^\mathrm{PL}$ peaks it
was not possible to extract reliable resonance traces due to
superpositions of different peaks.

The  T$_-^\mathrm{PL}$ peak gets resonant for $E_\mathrm{L}=1.341$
eV whereas T$_-$ has its resonance $E_\mathrm{L}=1.331$ eV. The
difference of these resonance energies of about 10 meV is in the
range of the difference in the energy shift between the
T$_-^\mathrm{PL}$ and the T$_-$ branch (cf.\ Figs.\
\ref{Resonanzmessung0T}, \ref{Resonanzmessung4_5T}, and
\ref{Uebersicht}). The full widths at half maximum (FWHMs) of the
resonance traces are about 14 meV for T$_-^\mathrm{PL}$ and 19.5 meV
for T$_-$ which is in the range of the linewidths obtained from the
nonresonant PL measurements (FWHM: 22 meV). The higher value for the
nonresonant case can be explained by additional multi-exciton
recombinations due to more charge carriers in the QDs because of the
nonresonant excitation.

The resonance energies for Q$_1$ and  Q$_2$ are about
$E_\mathrm{L}=1.344$ eV  and $E_\mathrm{L}=1.347$ eV, respectively,
in the range of the p$_h^-$-p$_e^-$ transition. Their different
maximum intensities have been discussed above. The difference in the
resonance laser energy between these excitations of about 3 meV
corresponds to the energy gap between the excitations Q$_1$ and
Q$_2$ as obtained from the measurements plotted in Figs.\
\ref{Resonanzmessung4_5T} and \ref{DispersionST}. The FWHMs of the
resonance traces are about 12 meV for each resonance which is a
smaller value than for the T$_-$ excitations. This can be explained
by the double-resonance condition necessary for these Raman
excitations as discussed above. Altogether, the resonance traces
support the proposed excitation scheme of a resonant PL together
with a resonant Raman process as sketched in Fig.\
\ref{DispersionST}(b).

The TS gets resonant for $E_\mathrm{L}=1.353$ eV which is in the range of the
p$_\mathrm{h}^+$-p$_\mathrm{e}^+$ transition in the single particle picture and which prompts us to assign the peak to a Raman transition from the T$_+$ to the S$_+$ state, as explained above. Surprisingly, the FWHM for this excitation of about 10 meV is even smaller than for the Q$_1$ or Q$_2$ excitations.

\subsubsection{The one-electron case}
In the following we discuss the one-electron case, i.\ e.,
measurements at a gate voltage of $V_\mathrm{g}=160$ mV where about
85 \% of the QDs contain $N=1$ electron. In the first instance one
would assume that the spectra in the this case should be much
simpler than in the two-electron case. However, it turns out that
the spectra are much more complex than anticipated. The reason are
strong polaronic effects which seem to be suppressed in the
two-electron case.

\begin{figure}
\includegraphics{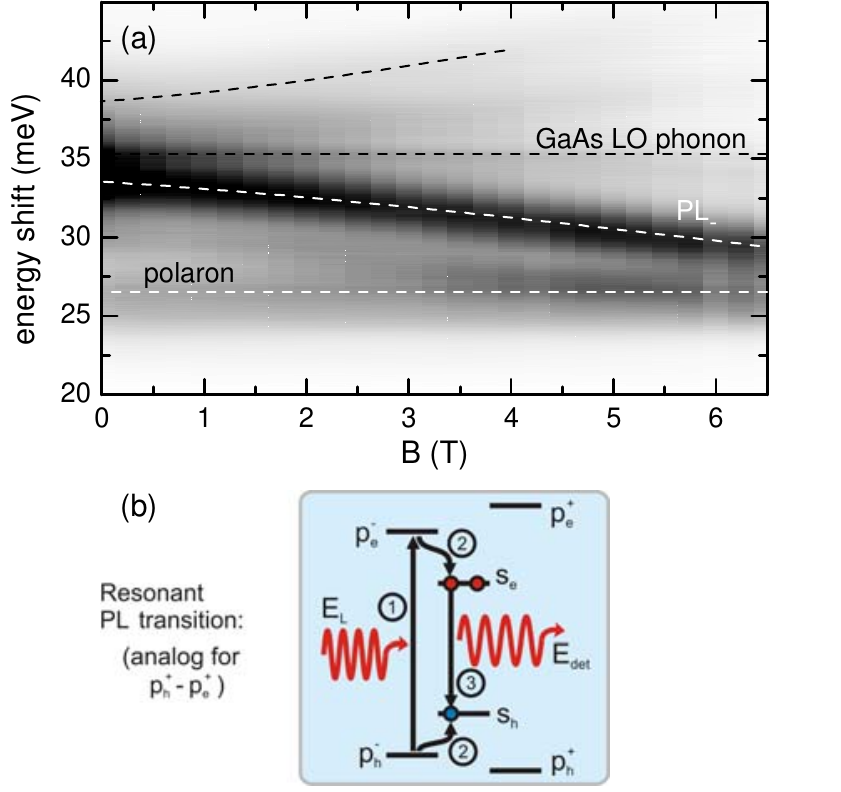}
\caption{\label{Dispersion1ePL}(a) Magnetic field dispersion of peaks resonantly excited with a
laser energy of $E_\mathrm{L2}=1.343$~eV. The branch labeled PL$_-$ is  assigned to a resonant PL process as sketched in (b) in a single-particle picture.}
\end{figure}

Figure \ref{Dispersion1ePL}(a) shows the magnetic field dispersion
of peaks occurring for an excitation laser energy of
$E_\mathrm{L2}=1.343$~eV. This figure is the one-electron analog of
the two-electron situation of Fig.\ \ref{ResonantPL}(b). The strong
branch with negative dispersion labeled as PL$_-$ is assigned to a
resonant PL process sketched in Fig.\ \ref{Dispersion1ePL}(b) in
the single-particle picture. In a first step, a resonant excitation
of a p$_h^-$-p$_e^-$ pair occurs. Then both, electron and hole,
quickly relax into their corresponding s state. The relaxation of
the electron is possible because, in contrast to the two-electron
case, the s shell is only half filled by only one electron. In a
third step, a radiative s$_e$-s$_h$ recombination occurs. The whole
process resembles the resonant PL process of the two-electron case
sketched in Fig.\ \ref{ResonantPL}(c). In a single-particle picture
the S$_-^\mathrm{PL}$ branch for the two-electron case and the
PL$_-$ branch should occur at the same energy shifts. The
fundamental difference between both processes is that in the
two-electron case, the final state splits up into singlet and
triplet states due to electron-electron interaction, whereas this is
not possible for the one-electron case. Consequently, equivalents of
the triplet branches do not appear in Fig.\ \ref{Dispersion1ePL}(a). The PL$_-$ branch occurs at slightly smaller energy shifts than
the S$_-^\mathrm{PL}$ branch, as has been reported for nonresonant
PL measurements on single charge-tunable InAs QDs.\cite{Findeis2001} The corresponding PL$_+$ branch involving a
p$_h^+$-p$_e^+$ transition is only allusively visible in Fig.\
\ref{Dispersion1ePL}(a) because of its larger resonance energy.
Besides the dispersive branches, we observe the non-dispersive GaAs
LO phonon branch at about 36 meV and a broader non-dispersive branch
at about 27 meV. The latter, we assign to a polaron peak as will be
discussed below.

\begin{figure}
\includegraphics{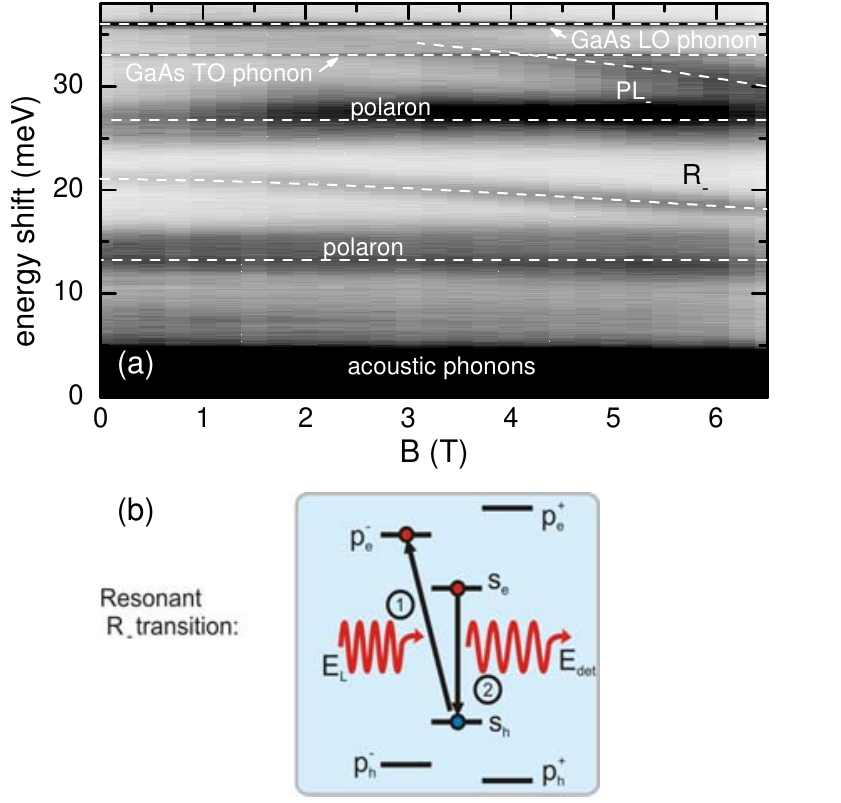}
\caption{\label{Dispersion1eRaman}(a) Magnetic field dispersion of peaks resonantly excited with a
laser energy of $E_\mathrm{L}=1.321$~eV. The branch labeled R$_-$ is
 assigned to a resonant Raman process as sketched in (b) in a single-particle picture.}
\end{figure}

Figure \ref{Dispersion1eRaman}(a) shows the magnetic field
dispersion of peaks occurring for an excitation laser energy of
$E_\mathrm{L}=1.321$~eV, smaller than $E_\mathrm{L2}$ used for the
spectra shown in Fig.\ \ref{Dispersion1ePL}(a). The observed peaks
occur at smaller energies. This figure is best compared to Fig.\
\ref{Raman194Wez}(a) of the two-electron case, however, for the
one-electron case in Fig.\ \ref{Dispersion1eRaman} we used a
slightly smaller excitation energy, since this delivers the best
overview of the dispersion of all observed peaks. For $B> 4$~T a
branch with negative dispersion occurs in the energy range of 30 meV
that can be assigned to the PL$_-$ peak discussed above. It is much
weaker than in Fig.\ \ref{Dispersion1ePL}(a) because the much lower
excitation energy selects a considerably smaller resonant
subensemble, as argued above. The sharp non-dispersive branches
at 36 and 33 meV are, respectively, the LO and TO phonon Raman
signals of the GaAs bulk material. Importantly, we assign the
dispersive branch in a range around 20 meV to a Raman scattering
process between the ground and the first excited state of the
one-electron QD. This branch is labeled R$_-$. Its two-step model
process is sketched in Fig.\ \ref{Dispersion1eRaman}(b): In the
first step a resonant s$_\mathrm{h}$-p$_\mathrm{e}^-$ excitation
occurs, followed by a radiative recombination between the electron
from the s$_\mathrm{e}$ state with the s$_\mathrm{h}$ hole.
Effectively, the electron of the one-electron QD is lifted from the
s$_\mathrm{e}$ ground  state to the p$_\mathrm{e}^-$ excited state.
Here, of course, the single particle wave functions of the initial
s$_\mathrm{e}$ and the final p$_\mathrm{e}$ state represent the
correct wave functions  without the need of constructing wave
functions by Slater determinants, as in the QD-helium case. We will
give a more detailed comparison of the single-particle excitation of
the one-electron QDs to the corresponding two-particle excitations
of the QD helium below. In Fig.\ \ref{Dispersion1eRaman}(a), the
excitation into the p$_\mathrm{e}^+$ state is not visible because of
a dominant nearly non-dispersive branch at about 27 meV, that in
outlines was already observed in Fig.\ \ref{Dispersion1ePL}(a).

The origin of the branch at 27 meV and of a similar branch at 13 meV
is not unambiguously clear. They might be assigned to polarons, i.\
e., strongly coupled modes between phonons and intersublevel
transitions in the QDs.\cite{Aslan2006} Interestingly, similar
branches do not occur for the two-electron case.  We are not aware
of any calculations for the one- or the many-body two-electron case
for the excitations in QDs coupled to phonons in the surrounding
media that explain strong polaron effects in one-electron QDs and
their strong suppression in two-electron QDs. So we have no good
explanation for this behavior. Nevertheless this behavior resembles
polaron effects on the cyclotron resonance in two-dimensional
electron systems, where they are completely screened at integer
filling factors, i.\ e., completely filled Landau levels (see, for
example, Refs.\ \onlinecite{Larsen1984}, \onlinecite{Peeters1985},
and references therein). This compares to the completely filled s
shell of the two-electron case in contrast to the not completely
filled s shell in the one-electron case. Such calculations would be
highly desirable and, we believe, of fundamental interest.

The last strong and yet undiscussed feature in Fig.\
\ref{Dispersion1eRaman}(a) is the region of strong intensity close
to 0 meV energy shift, i.\ e.\, close to the excitation laser energy
E$_\mathrm{L}$. We attribute this signals to a resonant PL process
with acoustic phonons involved.\cite{Favero2003,Urbaszek2004} The
laser light resonantly excites simultaneously a
s$_\mathrm{h}$-s$_\mathrm{e}$ transition together with acoustic
phonons. The detected light of the radiative
s$_\mathrm{e}$-s$_\mathrm{h}$ transition differs in energy from
$E_\mathrm{L}$ by the energy of the excited phonon.

\begin{figure}
\includegraphics{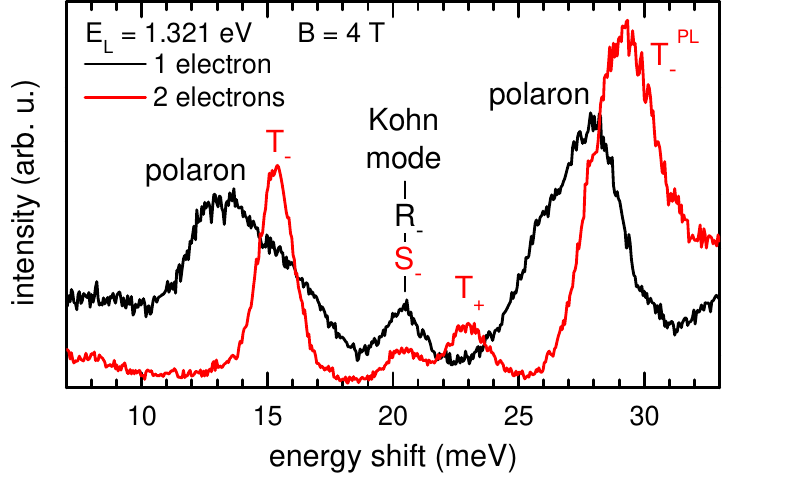}
\caption{\label{1vs2ecase}Depolarized spectra at a magnetic field
of $B=4$ T ($E_\mathrm{L}=1.321$ eV) for $N=1$ and $N=2$
electrons in the QDs are shown. 
The spectra for the one- and the two-electron case are
significantly different.}
\end{figure}

We now want to compare the Raman peaks of electronic excitations of
the one- and two-electron case. Figure \ref{1vs2ecase} shows
depolarized spectra at a magnetic field of $B=4$~T obtained for an
excitation laser energy of $E_\mathrm{L}=1.321$~ eV. The black (red)
spectrum stems from QDs containing $N=1$ ($N=2$) electrons. For QDs
charged with $N=1$ electron, the R$_-$ peak at about 21 meV is
visible, as well as the broad features around 13 meV and 27 meV that
we assign to polarons. On the other hand, for the QD helium case, the T$_-$,
S$_-$, T$_+$ and T$_-^\mathrm{PL}$ peaks, as discussed above, are
visible. The R$_-$ peak and the $S_-$ peak occur at the same energy
shift. This is a consequence of the generalized Kohn theorem
\cite{Maksym1990} which predicates that in QDs with a parabolic
potential the center-of-mass motion of the electrons is independent
of the number of electrons in the QDs. Thus, independent of the
number of electrons, the center-of-mass excitation occurs at the
quantization energy of the external potential, i.\ e.\, the single
particle quantization energy. The Kohn theorem holds also for
two-dimensional elliptical harmonic potentials,\cite{Li1991,Yip1991,Magnusdottir1999} as assumed for our QDs.
Particularly, in the QD helium, the transitions into the S$_-$ and
S$_+$ states correspond to the excitation of the center-of-mass
motion,\cite{merkt1991} thus they should occur at the same energy
of the single-particle excitations R$_-$ and R$_+$, as also can be
seen by comparing the results from theory in Fig.\ \ref{simu2}(c)
and (d). In the spectra of Fig.\ \ref{1vs2ecase}, the R$_+$ and
S$_+$ peaks are superimposed by, respectively, polaron and PL peaks,
thus only the R$_-$ and S$_-$ can be resolved and identified as the
Kohn mode. For the one-electron case, peaks corresponding to the
T$_-$, T$_+$ and T$_-^\mathrm{PL}$, as observed in the two-electron
case, are neither observed nor expected, since of course at least
two electrons are necessary to form triplet states. We also note
that we verified that no electronic Raman excitations occur in
uncharged QDs ($V_\mathrm{g}<60$ mV, cf.\ Fig.\ \ref{CV-Fit2})
because of the lack of electrons to be excited. Only PL peaks are
detected in this case.

\begin{figure}
\includegraphics{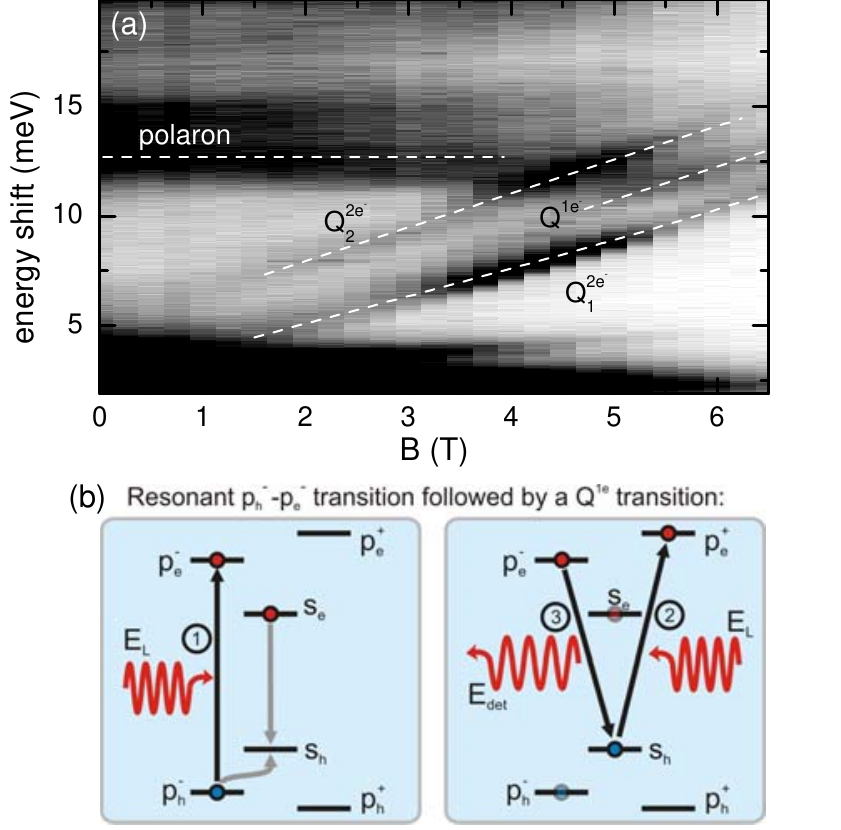}
\caption{\label{Dispersion1eQ}(a) Magnetic field dispersion of peaks resonantly excited with a
laser energy of $E_\mathrm{L}=1.351$~eV. The branches labeled Q$_{1/2}^{2\mathrm{e}}$ are assumed
to arise from the small fraction of QDs charged with two electrons and to correspond the the
Q$_{1/2}$ branches of Fig.\ \ref{DispersionST}(a). The branch labeled Q$^{1\mathrm{e}}$ is
assigned to a transition between the p$_\mathrm{e}^-$ and the p$_\mathrm{e}^+$ state of the
QDs charged by one electron. (b) Scheme of the doubly
resonant excitation process for Q$^{1\mathrm{e}}$ in a single-particle
picture. It consist of a resonant p$_\mathrm{h}^-$-p$_\mathrm{e}^-$ transition (left scheme) followed
by a resonant Raman process (right scheme).}
\end{figure}

Figure \ref{Dispersion1eQ}(a) shows the magnetic field dispersion
of peaks occurring with a small energy shift for an excitation laser
energy of $E_\mathrm{L}=1.351$~eV. This figure is best compared to
Fig.\ \ref{DispersionST}(a) of the two-electron case, however, for
the one-electron case in Fig.\ \ref{Dispersion1eQ} we used a
slightly larger excitation energy, since this delivers the best
overview of the dispersion of all observed peaks. Three dispersive
branches are observed. Two of them, labeled with
Q$_1^{2\mathrm{e^-}}$ and Q$_2^{2\mathrm{e^-}}$ occur exactly at the
same energies as the Q$_{1/2}$ branches in the two-electron case
(cf.\ Fig.\ \ref{DispersionST}). This is why we believe that they
arise from the small fraction of QDs which are not charged by $N=1$
but by $N=2$ electrons. From CV spectroscopy we can estimate the
fraction of doubly charged QDs to be only about 7\% (cf.\ Fig.\
\ref{CV-Fit2}). Until now we do not understand why only the Q
branches of the fraction of two-electron QDs occur, whereas no other
peaks peculiar for the QD helium case, i.\ e.\, the T$_-$, T$_+$, TS,
T$_-^\mathrm{PL}$, and T$_+^\mathrm{PL}$ peaks, are observed at the
gate voltage $V_\mathrm{g}=160$~mV. Thus, this assignment is not free
of ambiguity.

Interestingly, between the Q$_1^{2\mathrm{e^-}}$ and
Q$_2^{2\mathrm{e^-}}$ branches in Fig.\ \ref{Dispersion1eQ}(a), a
third branch labeled with Q$^{1\mathrm{e^-}}$ occurs. We assign this
branch to Raman transitions in singly charged QDs from electronic
states with $m=-1$ to states with $m=+1$. Figure \ref{Dispersion1eQ}(b) sketches a double-resonant excitation process in the
single-particle picture possibly underlying the Q$^{1\mathrm{e^-}}$
branch. Exciting with $E_\mathrm{L}$ resonantly  creates an
p$_\mathrm{e}^-$-p$_\mathrm{h}^-$ electron-hole pair. In the usual
resonant PL process, these carriers would quickly relax into their s
states and then finally radiatively recombine (gray arrows in the
left scheme). However, the Raman process proposed here has to happen
before the p$_\mathrm{e}^-$ electron relaxes. It is, in the
single-particle picture, independent of a possible relaxation of the
hole and a subsequent radiative recombination with the already
present s$_\mathrm{e}$ electron [this is why we depicted the hole
and the s$_\mathrm{e}$ electron transparent in the right scheme in
Fig.\ \ref{Dispersion1eQ}(b)]. The actual Raman process can be
considered as a two-step process: The first is a resonant
s$_\mathrm{h}$-p$_\mathrm{e}^+$ transition, the second is a
p$_\mathrm{e}^-$-s$_\mathrm{h}$ recombination. Thus, effectively,
the p$_\mathrm{e}^-$ electron is excited into to p$_\mathrm{e}^+$
state.

The process proposed for the Q$^{1\mathrm{e^-}}$ branch is similar
to the one described for the Q$_{1/2}$ branches of the QD helium.
Like the Q$^{2\mathrm{e^-}}_{1/2}$ branches, the Q$^{1\mathrm{e^-}}$
branch can only be observed for a certain range of magnetic fields,
for which the p$_\mathrm{h}^-$-p$_\mathrm{e}^-$ and
s$_\mathrm{h}$-p$_\mathrm{e}^+$ transition are of the same energy.
The weak intensity of this excitation compared to the counterparts
of the two-electron case can be explained by the
quick relaxation of the p$_\mathrm{e}^-$ electron into the s shell                                 before the actual Raman process takes place. This relaxation path is
blocked in the QD helium because of the completely filled s shell.
We note that for another sample we investigated, this peak is by a
factor of 2 stronger and thus can more clearly be identified. We
would expect that the Q$_2^{2\mathrm{e^-}}$ and Q$^{1\mathrm{e^-}}$
branches occur at the same energy shifts because of the generalized
Kohn theorem, that -- as described above -- predicts the
congruence between the S$_\pm$ and R$_\pm$ branches of the doubly-
and singly-charged QDs, respectively. However, the observed
deviations might arise through many-body effects of the hole created
in the first resonant transition and the s$_\mathrm{e}$ electron
[transparent in the right scheme of Fig.\ \ref{Dispersion1eQ}(b)]
or by polaronic effects.

\section{Conclusion}
In conclusion we report on resonant Raman and PL spectroscopy on
InGaAs QDs with an adjustable number of electrons. We find that the
spectra are significantly different for a charging of $N=1$ and for
$N=2$ electrons. For the QDs containing $N=2$ electrons resonant PL
and Raman scattering transitions from the ground into excited
triplet, the ortho He, and singlet, the para He, states are
observed. For the Raman transitions we demonstrate characteristic
polarization selection rules and the characteristic behavior of the
excitations for a transferred lateral wave vector. Also transitions
between excited triplet and singlet states as well as between
different angular momentum excited singlet or triplet states are
observed. For the one-electron case we also detect resonant PL
peaks, Raman scattering transitions from the ground to the first
excited state, and transitions between the excited states of
different angular momentum. Some of the signals obtained from
QDs with $N=1$ electron show strong polaronic effects.

\section*{Acknowledgements}
We gratefully thank W.\ Hansen, U.\ Merkt, C.\ Sch\"{u}ller, and Ch.\
Strelow for fruitful discussions. This work was supported by the
Deutsche Forschungsgemeinschaft via SFB 508 and GrK 1286.

\bibliography{Papersammlung}

\end{document}